\documentclass[twocolumn,twocolappendix]{aastex631}

\usepackage{graphicx}
\usepackage{comment}
\usepackage{natbib}
\usepackage{amsmath}
\usepackage{graphics}
\usepackage{enumitem}
\renewcommand{\baselinestretch}{0.96}

\newcommand{\elixer}{\texttt{ELiXer}}
\newcommand{\OII}{[\ion{O}{2}]}
\newcommand{\lya}{Ly$\alpha$}

\newcommand{\ndextotal}{1,642,390}
\newcommand{\ndexuse}{1,632,398}
\newcommand{\ncosmostotal}{32,557}
\newcommand{\ncosmosuse}{32,319}

\newcommand{\likely}{{\it Likely Real}}
\newcommand{\unlikely}{{\it Unlikely Real}}
\newcommand{\nlikely}{{4106}}
\newcommand{\nunlikely}{4375}
\newcommand{\nlikelylowsn}{{2301}}
\newcommand{\nunlikelylowsn}{{3914}}
\newcommand{\nlandnun}{8481}

\begin{document}

\title{ Enhancing Ly$\alpha$ Emitter Identification in HETDEX with a Convolutional Neural Network}


\correspondingauthor{} %
\email{shiro.mukae@shiseido.com}

\author[0000-0003-3823-8279]{Shiro Mukae}
\affiliation{Department of Astronomy, The University of Texas at Austin, 2515 Speedway Boulevard, Austin, TX 78712, USA}
\affiliation{MIRAI Technology Institute, Shiseido Co., Ltd., 1-2-11, Takashima, Nishi-ku, Yokohama, Kanagawa, 222-0011, Japan}

\author[0000-0002-2307-0146]{Erin Mentuch Cooper}
\affiliation{Department of Astronomy, The University of Texas at Austin, 2515 Speedway Boulevard, Austin, TX 78712, USA}
\affiliation{McDonald Observatory, The University of Texas at Austin, 2515 Speedway Boulevard, Austin, TX 78712, USA}

\author[0000-0002-8433-8185]{Karl Gebhardt}
\affiliation{Department of Astronomy, The University of Texas at Austin, 2515 Speedway Boulevard, Austin, TX 78712, USA}

\author[0000-0002-8925-9769]{Dustin Davis}
\affiliation{Department of Astronomy, The University of Texas at Austin, 2515 Speedway Boulevard, Austin, TX 78712, USA}

\author[0000-0002-1496-6514]{Lindsay R. House}
\affiliation{Data Science Institute, The University of Chicago, 5460 S University Ave, Chicago,
IL 60615, USA}
\affiliation{The SkAI Institute, 875 N. Michigan Ave., Suite 3500, Chicago, IL 60611}

\author[0000-0001-7066-1240]{Mahdi Qezlou}
\affiliation{Department of Astronomy, The University of Texas at Austin, 2515 Speedway Boulevard, Austin, TX 78712, USA}

\author[0000-0002-8984-0465]{Julian B.~Mu\~noz}
\affiliation{Department of Astronomy, The University of Texas at Austin, 2515 Speedway Boulevard, Austin, TX 78712, USA}

\author[0000-0002-6186-5476]{Shun Saito}
\affiliation{Institute for Multi-messenger Astrophysics and Cosmology, Department of Physics,
Missouri University of Science and Technology, 1315 N. Pine St., Rolla MO 65409, USA}
\affiliation{Kavli Institute for the Physics and Mathematics of the Universe (WPI), The University of Tokyo Institutes for Advanced Study (UTIAS), The University of Tokyo, Chiba 277-8583, Japan}

\author[0000-0003-2575-0652]{Daniel J. Farrow}
\affiliation{E. A. Milne Centre for Astrophysics
University of Hull, Cottingham Road, Hull, HU6 7RX, UK}
\affiliation{Centre of Excellence for Data Science,
Artificial Intelligence \& Modelling (DAIM),
University of Hull, Cottingham Road, Hull, HU6 7RX, UK}

\author[0000-0001-6842-2371]{Caryl Gronwall}
\affiliation{Department of Astronomy \& Astrophysics, The Pennsylvania
State University, University Park, PA 16802, USA}
\affiliation{Institute for Gravitation and the Cosmos, The Pennsylvania State University, University Park, PA 16802, USA}

\author[0000-0001-7240-7449]{Donald P. Schneider}
\affiliation{Department of Astronomy \& Astrophysics, The Pennsylvania State University, University Park, PA 16802, USA}
\affiliation{Institute for Gravitation and the Cosmos, The Pennsylvania State University, University Park, PA 16802, USA}

\author[0000-0003-1530-8713]{Eric Gawiser}
\affiliation{Physics and Astronomy Department, Rutgers, The State University, Piscataway, NJ 08854, USA}

\begin{abstract}
We present a deep learning framework to enhance the identification of Ly$\alpha$ emitters (LAEs) in the Hobby-Eberly Telescope Dark Energy Experiment (HETDEX), an untargeted spectroscopic survey of LAEs at $1.9 < z < 3.5$ without imaging pre-selection.
We primarily address the low signal-to-noise ratio (S/N) regime ($4.8 \leq \mathrm{S/N} \leq 5.5$), 
where LAE candidates suffer from substantial noise contamination.
To distinguish LAE candidates from artifacts and sky residuals, we employ a convolutional neural network (CNN) trained on two-dimensional spectral images of single emission lines.
The training sample is constructed from the HETDEX COSMOS catalog, with external validation from ancillary observations and our participatory science project, \textit{Dark Energy Explorers}. 
For small-format, low-resolution spectroscopic data, the model achieves a balanced accuracy, precision, and recall of $94.1\%$, $97.5\%$, and $97.5\%$, respectively, in the high-S/N regime ($\mathrm{S/N}>5.5$), and $85.1\%$, $78.2\%$, and $84.4\%$ in the low-S/N regime. 
Using HETDEX LAEs independently identified by DESI spectroscopy, the model recovers $99\%$ and $93\%$ of the high- and low-S/N LAEs, respectively.
Visual attribution indicates that the CNN attends to smooth, spatially extended central emission in true positives and to irregular or noisy features in true negatives.
Applied to the full HETDEX catalog, the CNN enables an S/N threshold down to 4.8 by suppressing spurious spikes across $z\sim 1.9$--$2.5$ in the redshift distribution. 
Our approach facilitates HETDEX cosmological analyses by mitigating false positives in galaxy clustering and highlights the value of domain-specific deep learning for refining low-S/N spectroscopic identification in untargeted surveys.
\end{abstract}

\keywords{
\href{http://astrothesaurus.org/uat/2171}{Galaxy spectroscopy(2171)}; 
\href{http://astrothesaurus.org/uat/978}{Ly-alpha galaxies(978)}; 
\href{http://astrothesaurus.org/uat/1938}{Convolutional neural networks (1938)}; 
\href{http://astrothesaurus.org/uat/582}{Galaxy classification systems (582)}; 
}

\section{Introduction} \label{sec:introduction}
Wide-field spectroscopic surveys are transforming observational cosmology by mapping the large-scale structure across vast spatial and redshift ranges, enabling measurements of baryon acoustic oscillations \citep[BAO;][]{Eisenstein2005,Cole2005},  providing precise constraints on cosmological parameters across cosmic time. 
Survey programs such as Sloan Digital Sky Survey \citep[SDSS:][]{York2000}, Baryon Oscillation Spectroscopic Survey \citep{Dawson2013}, and Extended Baryon Oscillation Spectroscopic Survey \citep{Dawson2016} over the past two decades, and more recently the Dark Energy Spectroscopic Instrument \citep[DESI;][]{desi2016, desi2022}, the \textit{Euclid} mission \citep{Mellier2025, Aussel2025}, and the Prime Focus Spectrograph \citep[PFS;][]{Takada2014, Tamura2024}, have been expanding the number of galaxy spectra from about $10^5$–$10^6$ in earlier surveys to 10$^7$ in ongoing and upcoming programs. 
These new programs push observations into deeper, higher-redshift regimes and increase diverse astronomical data, including imaging, photometry, and spectroscopy. Consequently, efficient and scalable analytical approaches are essential for extracting robust scientific insights.

A comparable large-scale galaxy survey is the Hobby-Eberly Telescope Dark Energy Experiment \citep[HETDEX;][]{Gebhardt2021, Hill2004}, an optical spectroscopic program aimed to identify over one million Lyman-$\alpha$ emitting galaxies (LAEs) at $1.9 < z < 3.5$ across $540~\mathrm{deg}^2$ of the sky, corresponding to a cosmic volume of $10.9~\mathrm{Gpc}^3$. One of HETDEX’s unique features is its untargeted survey design, in which spectra are obtained without prior imaging or target pre-selection.
HETDEX employs the Visible Integral-field Replicable Unit Spectrograph \citep[VIRUS;][]{Hill2021, Hill2002}, a wide-field, fiber-fed array of integral-field spectrographs mounted on the HET \citep{Ramsey1998}. The instrument comprises 78 integral field units (IFUs), each covering a $51\farcs \times 51\farcs$ region of the sky and feeding a pair of spectrographs through 448 fibers of $1\farcs5$ diameter. This configuration enables the simultaneous acquisition of $\sim35,000$ spectra per exposure. 
Over eight years of observations, HETDEX has collected more than 608 million fiber spectra across $87~\mathrm{deg}^2$ of sky, with a filling factor of 0.22, yielding 4.7 million detections of emission-line and continuum sources 
\citep[][in preparation; hereafter MC26]{MentuchCooper2026a}
The observed LAEs serve as tracers of the large-scale structure \citep{Ouchi2020}, and are used to constrain the Hubble parameter $H(z)$ and angular diameter distance $D_A(z)$ to percent-level precision \citep{Gebhardt2021}. 

\begin{figure*}[!t]
\centering
    \includegraphics[trim=10 20 80 120, clip, width=1.0\textwidth]{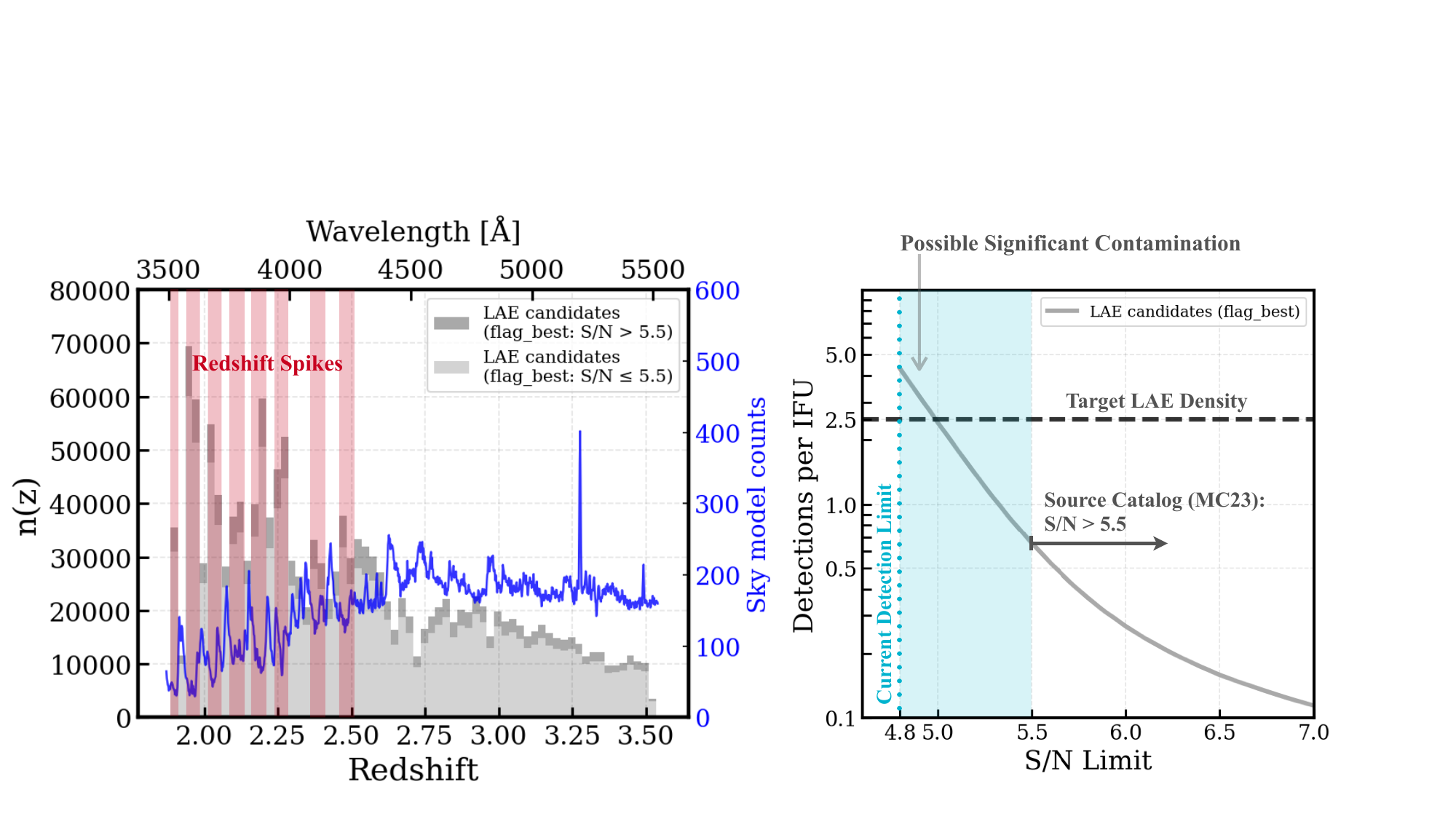}
    \caption{
    {\bf Redshift distribution and cumulative number of HETDEX LAE candidates as a function of S/N limits.} 
    {\bf Left:} Redshift histograms of LAE candidates selected with data quality flags (MC26). The bin width is $\Delta z = 0.025$. The dark and light gray bars indicate the LAE candidates with $\mathrm{S/N} > 5.5$ and $\mathrm{S/N} \leq 5.5$, respectively. 
    The blue spectrum depicts the sky model in the HETDEX survey \citep{Gebhardt2021}.
    The red-shaded regions indicate spurious redshift spikes, likely due to false detections, 
    caused by reduced VIRUS sensitivity or dips in the sky model where read noise dominates and sky subtraction residuals can appear as spurious emission-like features. 
    {\bf Right:} 
    The solid curve shows the cumulative number of LAE candidates per VIRUS IFU as a function of the S/N threshold, considering only the data quality flag-based selection.  
    The horizontal arrow marks the current identification threshold adopted in HETDEX public source catalog of \citet{MentuchCooper2023}. 
    The vertical dotted line indicates the current HETDEX detection limit ($\mathrm{S/N}=4.8$). 
    The dashed line denotes the target LAE density in the HETDEX survey \citep{Gebhardt2021}. 
    The vertical arrow implies possible significant contamination, where the density of LAE candidates 
    exceeds the expected target density, introducing contamination from spurious detections.}
    The cyan region highlights the S/N range in which this study aims to extend the identification threshold and distinguish LAEs from spurious detections.
    \label{fig:flag_best}
\end{figure*}

A key challenge in HETDEX is the identification of LAEs in the low signal-to-noise ratio (S/N) regime.
Figure~\ref{fig:flag_best} shows the redshift distribution and the cumulative number of LAE candidates per VIRUS IFU as a function of S/N limit. 
The left panel presents histograms of LAE candidates 
selected by the HETDEX Emission Line eXplorer software tool \citep[ELiXer;][]{Davis2023} 
after using a composite of data-quality flags (\texttt{flag\_best}$=1$) 
to exclude artifacts, bad fibers, bad pixels, and contamination from satellites and meteors (MC26). In the right panel, the solid curve indicates cumulative number of the LAE candidates. 
The HETDEX public source catalog \citep{MentuchCooper2023} adopts a conservative identification threshold of $\mathrm{S/N} > 5.5$ for emission lines, combined with the quality flags to reduce contamination. While effective at minimizing spurious detections, this criterion also excludes a substantial number of LAEs, lowering the source density below the target of $2.5$ LAEs per IFU and limiting the statistical power for cosmological analyses such as BAO measurements.
The target LAE number density is estimated by integrating published LAE luminosity functions \citep{Gronwall2007, Ouchi2008} down to the HETDEX flux limit for nominal observing conditions
(a base exposure time of 18 min split into three 6 min dithers).
Extending the current selection threshold from $\mathrm{S/N} = 5.5$ into the lower S/N regime, down to the current detection limit of $4.8$, 
would exceed the target density at the cost of introducing significant contamination from spurious detections. 

At lower redshifts ($z \lesssim 2.5$), where the sensitivity of  VIRUS decreases toward shorter wavelengths ($\lambda \lesssim 4300$\,\AA; \citealt{Hill2021}), prominent spurious spikes appear,  as highlighted in the red-shaded regions of Figure~\ref{fig:flag_best}.
The number of LAE candidates fluctuates by up to a factor of $\sim2$ between adjacent $\Delta z = 0.025$ bins, even though the bright end of \lya\ luminosity functions evolve monotonically across $z \sim 2$–$3$ \citep{Nagaraj2025, Umeda2025}, which is the luminosity regime probed by HETDEX \citep{Zhang2021}. 
These spikes are likely caused by contamination from false detections, which are not randomly distributed in redshift space and can mimic clustering signals, potentially biasing the LAE clustering analysis.
The redshift spikes may reflect the reduced VIRUS sensitivity and arise from sharp dips in the HETDEX sky model \citep{Gebhardt2021}, where the background sky level is low and read noise dominates.
These effects cause an underestimation of the noise level at these wavelengths. As a result, sky-subtraction residuals can mimic emission-line signals, producing numerous false detections.
Such false positives are not fully accounted by the current noise model or quality flags, making it difficult to distinguish true emission lines from artifacts and residuals.
Given the more than one million LAE candidates in the survey, visual inspection is labor-intensive and particularly challenging in the low-S/N regime, motivating the development of automated methods to suppress contamination and recover LAEs effectively.

Deep learning has emerged as a promising tool for automated source identification and classification in large astronomical datasets. Applications in galaxy surveys have demonstrated its effectiveness in identifying faint galaxies \citep[][]{Huertas-Company2018, Ono2021}, characterizing galaxy spectra \citep{Melchior2023, Hahn2025, TardugnoPoleo2023}, and detecting artifacts or poor-quality exposures \citep[][]{Zhang2020, Chang2021, Tanoglidis2022, Luo2025}. These methods are particularly valuable for bridging the gap between low-level data processing and high-level scientific interpretation, addressing the challenges posed by the rapidly growing data volumes from current and next-generation surveys \citep[][]{Siudek2025, Parker2024}.

In this paper, we present a deep learning application for identifying LAEs and filtering noise contaminants in the HETDEX untargeted spectroscopy, with the goal of extending the S/N threshold to lower limits and increasing the number density of genuine LAEs.
We develop a  convolutional neural network \citep[CNN;][]{Cun1989}, designed to extract morphological features through hierarchical convolutional operations, and address following key challenges: 
(i) Low-S/N source classification – developing a CNN model that distinguishes \lya\ emission from spurious features using two-dimensional (2D) spectral images of single emission lines on the detector plane;
(ii) Data scale and resolution — tailoring the CNN to operate effectively on small-format and low-resolution 2D spectral images;
(iii) Limited label availability – constructing a reliable training set utilizing external validation data and contributions from our participatory science project; and 
(iv) Interpretability — integrating attribution-based visualization methods to highlight the key input features driving the model’s predictions \citep{Feng2024, Marcinkevis2023}.
These approaches improve both the recovery and reliability of spectroscopic LAE identification in the low-S/N regime, complementing machine learning methods pioneered in early HETDEX data \citep{House2024, Sakai2021},
and establish a methodological basis for applying deep learning to future large-scale spectroscopic surveys.

This paper is organized as follows. Section~\ref{sec:data} describes the HETDEX catalogs and spectra used in this study. Section~\ref{sec:methods} outlines our methodology, including  training data preparation, CNN construction and training, and the visual attribution technique. Section~\ref{sec:results and discussions} presents the results and discussion, including CNN classification performance, attribution visualizations, and the application to the full HETDEX catalog. 
Finally, Section~\ref{sec:conclusions} summarizes our conclusions. Throughout this paper, we adopt the flat $\Lambda$CDM cosmology with $H_0=67.7\,\mathrm{km}\,\mathrm{s}^{-1}\,\mathrm{Mpc}^{-1}$ and $\Omega_{\mathrm{m},0}=0.31$ measured by \citet{Planck2018}. 
All magnitudes are expressed in the AB system \citep{oke1983}. 

\begin{figure*}[!t]
\centering
    \includegraphics[trim= 10 0 0 220, clip, width=1.0\textwidth]{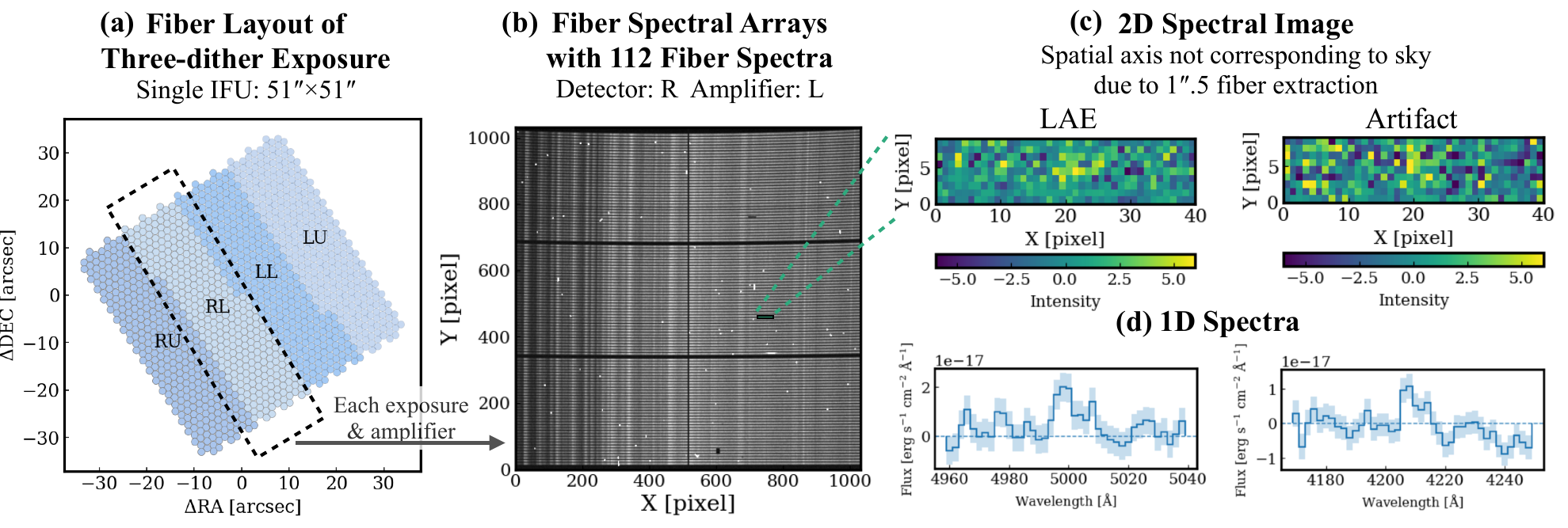}
    \caption{
    {\bf Schematic illustration of 
    (a) fiber layout, (b) fiber spectral arrays, (c) 2D spectral images, and (d) 1D spectra.}
    {\bf (a):} 
    Fiber layout of a three-dither HETDEX observation for a single $51\farcs \times 51\farcs$ IFU. 
    VIRUS consists of 78 IFUs in total. Each filled circle marks the on-sky footprint of a $1\farcs5$-diameter fiber, showing the positions sampled in the three dithered exposures. 
    The different shades denote the four amplifier channels within the IFU. 
    For each exposure and amplifier channel, 112 fiber spectra are obtained. 
    The configuration of the IFU and its amplifiers is detailed in Figure~2 of \citet{MentuchCooper2023} and Figure~4 of \citet{Gebhardt2021}.        
    {\bf (b):} 
    Raw image of fiber spectral arrays associated with an IFU and one of its four amplifiers (IFU 076, amplifier ``RL''). The 112 fiber spectra are encoded on the 1032~pixel~$\times$~1032~pixel detector. Each 2D spectral image is produced by summing the sky-subtracted spectral arrays for fibers centered on the source position, using PSF-based weights.
     {\bf (c):} 
    2D spectral images of an LAE candidate and an artifact with $4.8 \leq \mathrm{S/N} \leq 5.5$ shown as cutouts of 9 cross-dispersion pixels by 40 dispersion pixels. The cross-dispersion axis does not correspond to sky coordinates because each spectral image is not spatially resolved within the fiber's $1\farcs5$ diameter.
    The LAE candidate is drawn from the ODIN-confirmed LAEs \citep{Firestone2024}, whereas the artifact is taken from the HETDEX emission-line detections with a conservative vote-fraction of \texttt{dee\_prob} $\leq 0.1$ (classified as "throwback” false detections) in {\it Dark Energy Explorers}.    
    {\bf (d):}
    Corresponding 1D spectra of the sources in panel (c). While the LAE and artifact appear similar  in 1D, they can be discerned when viewed in 2D.
    }
    \label{fig:array_2d_1d}
\end{figure*}

\section{Data} \label{sec:data}
We construct a CNN model using HETDEX 2D spectral images of \lya\ emission-line candidates from the HETDEX COSMOS catalog and subsequently apply it to the full HETDEX dataset (hereafter the HETDEX DEX catalog)
to distinguish astronomical sources from artifacts and sky residuals.
Details of the HETDEX catalogs and the 2D spectral images are provided in Sections \ref{subsec:hc} and \ref{subsec:hs}, respectively.

\subsection{HETDEX Catalogs} \label{subsec:hc}
\subsubsection{HETDEX Emission-Line Data} \label{subsubsec:el}
The \lya\ emission line candidates used in this study are drawn from the HETDEX Internal Data Release 5.0.1 (HDR5; released internally on 2025 May 31), which encompasses HETDEX observations conducted from January 1, 2017, through July 31, 2024. The HDR5 contains approximately $1.6$ million LAE candidates with $\mathrm{S/N} \geq 4.8$ and  $\chi^2 < 2.5$, based on Gaussian fit in spectral axis. 
The S/N of each detected line is computed as the ratio of the fitted emission-line flux to the local noise level by grid searches of emission-line candidates in all spectral and spatial resolution elements \citep[see Section 7 in][]{Gebhardt2021}. 
The HETDEX observations reach 50\% completeness at a line flux of \( \sim 1.1 \times 10^{-16}~\mathrm{erg~s^{-1}~cm^{-2}} \), corresponding to Ly\(\alpha\) luminosities as low as \( \sim 1.0 \times 10^{42.8}~\mathrm{erg~s^{-1}} \) at \( z \sim 2.7 \).
This 50\% completeness flux limit is wavelength dependent and is derived from source injection simulations that add mock \lya\ lines to actual observed data, propagate element-by-element noise models, and measure the recovery fraction with the HETDEX detection pipeline. 
Despite the wavelength dependence, the recovery fractions are nearly identical across the HETDEX science-verification fields at any fixed wavelength; the completeness curve is therefore obtained by averaging the field-by-field curves within 
$\sim270$Å wavelength bins \citep[see Fig.~28 of][]{Gebhardt2021}.

Because the VIRUS integral-field spectrographs have narrow wavelength coverage ($3500-5500 {\rm \AA}$) and low spectral resolution ($R\sim800$), LAEs at $1.9 < z < 3.5$ and \OII\ emitters at $z < 0.5$ are typically observed as single emission-line sources. Since the spectral resolution of VIRUS is insufficient to resolve either the asymmetry of the Ly$\alpha$ emission-line profiles or the \OII\ $\lambda3727$ doublet, 
LAEs and [O\,II] emitters cannot be readily distinguished \citep[see Figure~1 of][]{Leung2017}.
These high-redshift \lya\ and low-redshift \OII\ emission lines are classified in the HETDEX ELiXer software tool \citep{Davis2023}. ELiXer is based on the Bayesian analysis of line luminosity and photometric equivalent width derived from ancillary broadband imaging data \citep{Leung2017}, which is further developed into an LAE probability estimator through the implementation of \citet{Farrow2021} to evaluate the likelihood of Ly$\alpha$ relative to \OII\ based on line luminosity functions and equivalent-width distributions \citep{Gronwall2007, Ciardullo2012, Ciardullo2013}. ELiXer expands on these analyses of emission line equivalent widths by additionally incorporating multiple metrics from the HETDEX spectra and archival photometric imaging and catalogs. This approach provides reliable source classifications. 
{\color{blue} In the faint regime (${g}_{\rm mag} > 22$), misclassifications are found but constitute only a small fraction of the sources, where the continuum emission from \OII\ emitters is typically undetectable.}

The main objective of this study is to distinguish astronomical emission lines from artifacts and sky residuals. ELiXer also incorporates disqualification checks to remove spurious detections caused by artifacts or poor data quality \citep{Davis2023}. As a whole, this study assesses whether the emission line evaluated by ELiXer, in combination with data quality flags (MC26) that are used in the HETDEX source catalog generation, represents a true astrophysical source.
A comprehensive classification of LAE and \OII\ emitters is undertaken separately by alternative machine learning techniques, including a Random Forest classifier (MC26).

\subsubsection{HETDEX DEX and COSMOS Catalogs} \label{subsec:hdcc}
This study uses two \lya\ emission-line catalogs from HDR5, the HETDEX DEX and COSMOS catalogs, whose basic properties are summarized in Table~\ref{tab:hetdex_catalogs}.

The DEX catalog includes \ndextotal\ LAE candidates at $1.9 < z < 3.5$ across the survey area of $86.67~\mathrm{deg}^2$. Because the HETDEX observations are tiled and non-contiguous, the VIRUS focal plane has filling factor of $22\%$ \citep{Gebhardt2021, Hill2021}.
This catalog providing source coordinates, redshifts, S/N, source classifications as either ‘LAE’ or ‘AGN’, and SDSS $g$-band magnitudes (${g}_{\rm mag}$) measured from HETDEX spectra.
Through emission-line detection and catalog production, data quality filtering with the \texttt{flag\_best} indicator excluded obvious artifacts and unreliable detections (e.g., those caused by bad fibers, defective pixels, satellite trails, or meteor streaks). Nevertheless, the DEX catalog still contains ambiguous cases where distinguishing between astrophysical signals from spurious ones remains difficult. 

To address this issue, we employ the HETDEX COSMOS catalog, a subset of the DEX catalog augmented with multi-wavelength observations and additional validation data in the COSMOS field. 
The ancillary data used in this study are described in Section~\ref{subsec:lc}. 
The COSMOS catalog is designed to characterize spurious detections that remain after quality filtering and to facilitate the development of machine learning models for their classification. This catalog contains \ncosmostotal\ LAE candidates over the $2.26~\mathrm{deg}^2$ COSMOS field. 

In this study, Active Galactic Nuclei (AGN) candidates are excluded from both the DEX and COSMOS catalogs. We remove AGN candidates with the HDR5 \texttt{flag\_agn} indicator, based on the AGN catalog compiled by \citet{Liu2022, Liu2025}. This catalog includes 15,940 sources across the redshift range $z = 0.1$--$4.6$, cross-matched with the SDSS Quasar Catalog DR16Q \citep{Lyke2020}, as well as AGN candidates identified by spectral features such as single broad emission lines with a rest-frame full width at half maximum (FWHM) greater than 1200~km~s$^{-1}$, or line pairs of \lya\ and highly ionized lines (e.g., C\,\textsc{iv}~$\lambda1549$).
Such AGN sources could bias the model toward AGN-like spectral characteristics, thereby degrading its performance in distinguishing noisy LAE candidates from spurious detections, especially in the low-S/N regime.
The resulting source counts are \ndexuse\ for the HETDEX DEX catalog and \ncosmosuse\ for the COSMOS catalog, as summarized in Table~\ref{tab:hetdex_catalogs}.

\begin{deluxetable*}{lccccc}[t]
\tabletypesize{\small}
\tablecaption{Summary of HETDEX \lya\ Emission Line Catalogs  
\label{tab:hetdex_catalogs}}
\tablehead{
\colhead{Catalog} & \colhead{Fields} & \colhead{Total Area} & \colhead{Number Count} & \colhead{Number Count (after AGN removal)}}
\startdata
HETDEX DEX & DEX-Spring, DEX-Fall, COSMOS,  & 86.67 deg$^2$ & \ndextotal\ & \ndexuse\ \\
 & NEP, SSA22, and GOODS-N & & & \\
HETDEX COSMOS & COSMOS & 2.26 deg$^2$ & \ncosmostotal\ & \ncosmosuse\ \\
\enddata
\tablecomments{The data in this paper are based on HETDEX Internal Data Release 5.0.1.}
\end{deluxetable*}
\begin{figure*}[t]
\centering
    \includegraphics[trim=30 0 170 330, clip, width=1.0\textwidth]{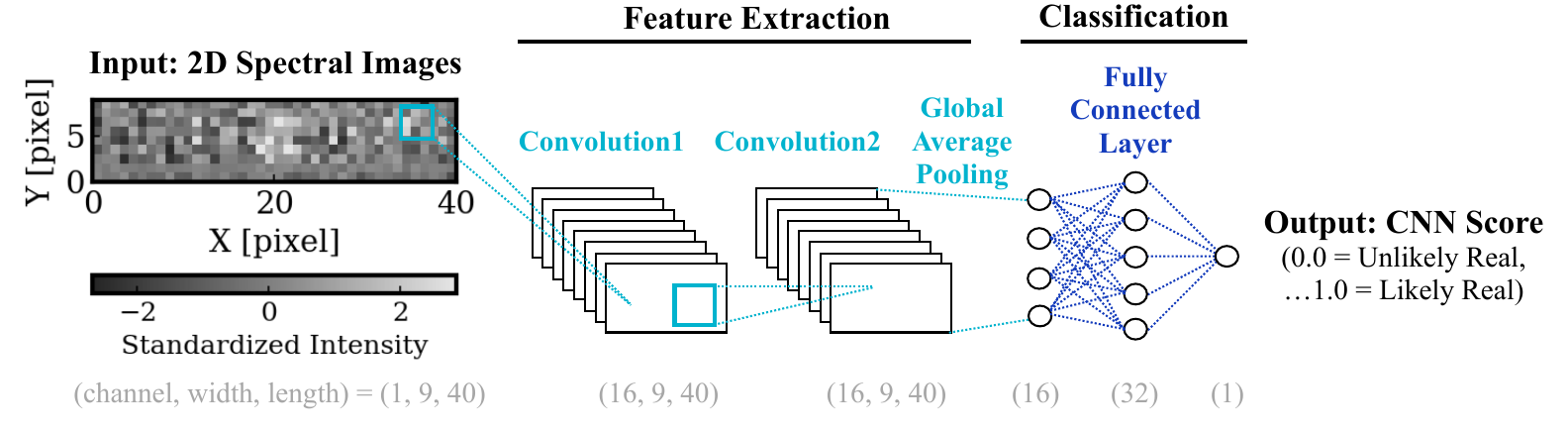}
    \caption{
    {\bf Schematic overview of the  CNN architecture employed in this work.}
    The model takes a 2D spectral image (9~$\times$~40 pixels) as input and outputs a confidence score between 0.0 and 1.0.
    Feature extraction is performed using two blocks of convolutional layers, followed by global average pooling and classification with two fully connected layers. 
    The final CNN score represents the model's confidence in the presence of an emission line. 
    The model architecture, including the filter sizes, number of channels, batch normalization, and dropout, is listed in Table~\ref{tab:model_architecture}.
    }
    \label{fig:model_architectures}
\end{figure*}

\subsection{HETDEX Spectral Images} \label{subsec:hs}
The HETDEX 2D spectral images are constructed from compilations of fiber spectral arrays obtained with the VIRUS integral-field spectrographs. In Figure~\ref{fig:array_2d_1d}, panels (a), (b), and (c) illustrate the fiber layout, fiber spectral arrays, and 2D spectral images, respectively. 
During a typical HETDEX observation with three dithered exposures,
the light collected by each IFU and its amplifier channel is dispersed by a spectrograph, and a total of 112 fiber spectra are recorded on the 1032~pixel~$\times$~1032~pixel detecter.
The 2D spectral image of each LAE candidate is created by summing the sky-subtracted fiber spectral arrays for fibers centered on the source position. 
As shown in Figure~2 of \citet{MentuchCooper2023}, multiple fibers whose sky coordinate lie within a $3\farcs5$ circular aperture contribute to each summed spectrum, with fluxes weighted by a point-spread function (PSF) model. 
The weighting follows the optimal-extraction algorithm of \citet{Horne1986}, assuming a symmetric two-dimensional Moffat profile ($\beta = 3.0$; \citealt{Moffat1969}) derived from stellar sources in the same exposure. 
In the HETDEX pipeline, each emission-line detection utilizes $\sim$20 fibers combined using PSF-based weights, whereas the CNN input 2D spectral image is constructed by summing the 2D spectral images of the four fibers with the highest PSF weights, as most of the source flux is contained within these fibers \citep{Davis2023}.

The 2D spectral images are taken from the HDR5 data using the \texttt{hetdex\_tools.get\_spec2D} function from the \texttt{hetdex\_api} package\footnote{\url{https://github.com/HETDEX/hetdex_api}}. 
The spectral image (before cutout) covers a wavelength range of $3500$–$5500~{\rm \AA}$ along the dispersion axis, with 1032 pixels and a resolving power of $R \sim 800$ (corresponding to a spectral resolution of $\sim400~{\rm km~s^{-1}}$). For each emission-line, we extract a spectral window of $\pm 40 {\rm \AA}$ (approximately $\pm2000\ \text{km s}^{-1}$) centered on the line, following the visual diagnostics of \cite{House2023} and \cite{Davis2023}. Since most LAE candidates,  typically low-mass galaxies, show no discernible continuum or additional emission lines in the HETDEX spectra due to limited sensitivity \citep{ChvezOrtiz2023}, including the entire spectrum would only introduce additional noise and hinder training. By restricting the input to the emission-line region, we optimize the CNN to focus on the relevant and physically meaningful features.

The resulting 2D spectral images have a shape of $(1, 9, 40)$, corresponding to one intensity channel, nine spatial (cross-dispersion) pixels, and forty spectral (dispersion) pixels, respectively. 
The spatial axis on the detector does not correspond to sky coordinates since each spectral image is extracted from a $1\farcs5$-diameter fiber and is not spatially resolved within that diameter.
Importantly, combination of the dispersion and cross-dispersion information allows the CNN to effectively recognize characteristic morphologies of emission lines and artifacts that would otherwise be indistinguishable in one-dimensional (1D) spectra. 
Because emission lines and false detections can exhibit similar 1D profiles, classification based solely on 1D spectra is inherently difficult.
Panel (d) of Figure~\ref{fig:array_2d_1d} shows examples where LAEs and artifacts appear similar in 1D but are distinguished when viewed in 2D.
We therefore employ 2D spectral images as CNN inputs to better capture emission-line structures.

To prepare the data for model training, we standardize the intensities of the 2D spectral images. 
An example input is shown in Figure~\ref{fig:model_architectures}. Details of the training data and model setup are described in the following Section~\ref{subsec:lc}.

\section{Methods} \label{sec:methods}
This section describes the development of the CNN model.
Section~\ref{subsec:lc} presents the construction of the training sample.
Section~\ref{subsec:cnn} provides the CNN architecture, model training, and performance evaluation methods. 
Section~\ref{subsec:vat} outlines the visual attribution technique used to enhance model interpretability.

\subsection{Training Sample Construction} \label{subsec:lc}
The construction of representative training samples for \lya\ emission lines and spurious detections is one of the most challenging aspects of this study. 
In the HETDEX survey, which employs untargeted spectroscopy, a single emission line in the low-S/N regime provides insufficient information to confidently distinguish genuine astronomical sources from artifacts.
To verify the sources responsible for the detected lines, HETDEX relies on ancillary data. Broadband imaging in the survey fields assists in identifying continuum counterparts \citep{Davis2023}. However, LAEs are typically low-mass stellar systems with faint rest-UV continua due to their low star-formation rates, so their counterparts may remain undetected in the ancillary imaging data.

Given these challenges, we construct the training sample for our CNN model as follows. We select high-confidence \lya\ emission lines and artifacts by leveraging ancillary multi-wavelength data and additional verification diagnostics from the HETDEX COSMOS catalog. These include (i) imaging and spectroscopic surveys, (ii) HETDEX repeat observations, and (iii) classifications from our participatory science project. The specific selection criteria for each category are described below.

\paragraph{\likely}  
This category refers to sources considered likely to be galaxies, selected if any of the following criteria are satisfied:
\begin{itemize}
  \item Detections whose \lya-emission redshifts are confirmed by external surveys and compilations. These include narrow-band surveys such as SILVERRUSH \citep[$z=2.2$ and $3.3$;][]{Kikuta2023} and ODIN \citep[$z=2.4$ and $3.1$;][]{Firestone2024}, and narrow- and medium-band surveys of SC4K \citep[$z\sim 2.2\ –\ 3.4$;][]{Sobral2018}, 
  as well as by spectroscopic \citep{Khostovan2026} or photometric \citep{Weaver2022} redshift compilations in the COSMOS field. Detections are cross-matched to external surveys using a 1\arcsec\ matching radius. 
  For spectroscopic samples, matches are accepted when $Delta z /(1+z) < 0.02$. For photometric redshifts, agreement is required at $\Delta z /(1+z) < 0.025$.
  For narrow-band surveys, a detection is considered a match if the observed HETDEX \lya\ wavelength falls within $\pm$FWHM the central wavelength of the corresponding narrow-band filter curve.
  \item Sources verified by visual inspection of repeat detections in multiple HETDEX observation shots \citep{Gebhardt2021}. The COSMOS field includes regions with multiple HETDEX exposures. We visually confirm using \elixer\ reports \citep{Davis2023} that the repeat HETDEX detections are not an artifact.
  \item Sources classified as ``keep'' in the NASA Zooniverse\footnote{\url{https://www.zooniverse.org}} participatory science project, {\it Dark Energy Explorers} 
  \citep[DEE;][]{House2023}\footnote{\url{https://www.zooniverse.org/projects/erinmc/dark-energy-explorers}} with a vote fraction  \texttt{dee\_prob} $\geq 0.7$.  
  In the DEE workflow ``Fishing for Signal in a Sea of Noise'', the primary goal is to further reduce false detections in the HETDEX source catalog\footnote{In the HDR5 data, this DEE workflow is applied to sources with $\mathrm{S/N} > 5.5$, except in the COSMOS fields, where classifications are also extended to the lower-S/N regime ($4.8 \leq \mathrm{S/N} \leq 5.5$). The DEE workflow for this lower-S/N regime is still in progress. This limited coverage motivates our combined use of CNN and DEE to reduce false detections across the HETDEX full sample.}.
  The parameter \texttt{dee\_prob} is defined for each source as the fraction of ``keep'' votes among all binary classifications 
  (``keep'' for likely LAE vs.``throwback'' for false detection), contributed by more than ten participants. Thus, \texttt{dee\_prob} close to 1.0 suggests broad consensus that the source is likely an LAE, whereas a value near 0.0 indicates that it is false detection. While the classification accuracy of \texttt{dee\_prob} largely agree with those from the HETDEX astronomer team for sources with $\mathrm{S/N} > 5.5$ \citep{House2023}, the lower-S/N regime ($4.8 \leq \mathrm{S/N} \leq 5.5$) remains under verification. To address this, we adopt a moderately conservative threshold of \texttt{dee\_prob} $\geq 0.7$ (i.e., at least $70\%$ of participants classified the source as an LAE), as our visual inspection indicates that this cut effectively removes false detections, while retaining a sufficient number of low-S/N sources for training.
\end{itemize}

\paragraph{\unlikely}  
This category refers to sources identified as spurious detections, selected if both of the following criteria are satisfied:
\begin{itemize}
\item Sources classified as ``throwback'' in the DEE participatory science project, with a conservative vote fraction threshold, \texttt{dee\_prob} $\leq 0.1$ (i.e., at most $10\%$ of participants classified the source as an LAE).
\item Sources not overlapping with ancillary datasets used to define the {\likely} category.
\end{itemize}

The constructed training sample contains a total of \nlandnun\ sources, 
comprising \nlikely\ \likely\ and \nunlikely\ \unlikely\ objects, 
corresponding to an overall class balance of 48/52.
The class counts and their breakdown by S/N regime are summarized in
Tables~\ref{tab:Likely_Real} and \ref{tab:Unlikely_Real}.
In the low-S/N regime ($4.8 \leq \mathrm{S/N} \leq 5.5$), the sample contains 
\nlikelylowsn\ \likely\ and \nunlikelylowsn\ \unlikely\ objects (37/63),
whereas in the high-S/N regime  ($\mathrm{S/N} > 5.5$), it contains 1805 Likely Real and 461 Unlikely Real objects (80/20). 
The constructed training sample exhibits imbalances between \likely\ and \unlikely\ and between the low- and high-S/N regimes, arising from the combination of selection thresholds used to define the labels. 
To mitigate potential biases in model training, 
we stratify the data by both label and S/N regime when constructing the training/validation/test splits and the cross-validation folds (Section~\ref{subsec:mt}).

\begin{figure*}[p]
\centering
    \includegraphics[trim=0 0 200 0, clip, width=0.8\textwidth]{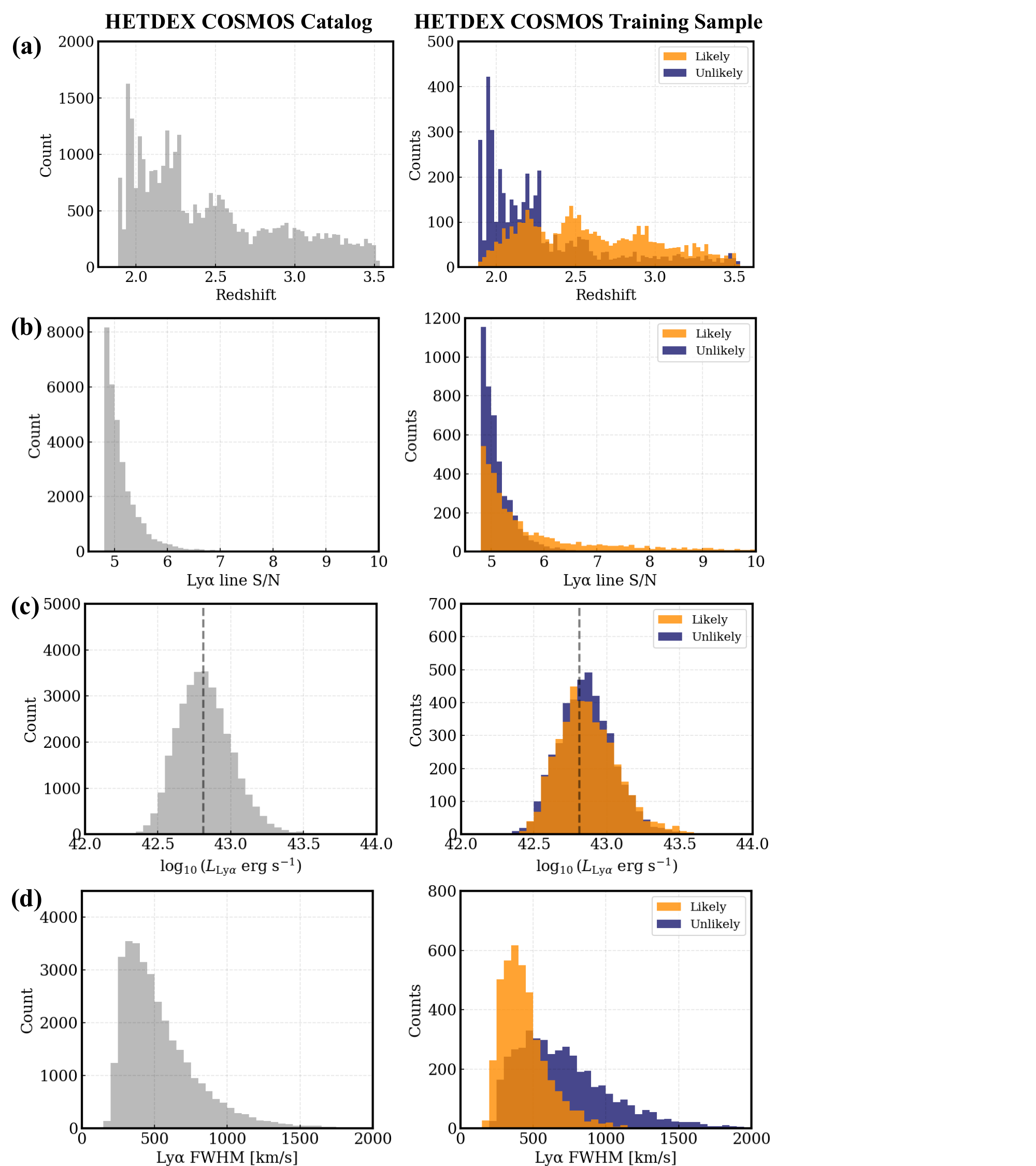}
    \caption{
    {\bf Histograms for HETDEX COSMOS LAE candidates: (a) Redshift, (b) \lya\ line S/N, 
    (c) \lya\ luminosity, and (d) \lya\ FWHM from Gaussian fits.}
    {\bf Left:} Distribution of \ncosmosuse\ LAE candidates in the HETDEX COSMOS catalog.
    {\bf Right:} Subsample of the HETDEX COSMOS catalog used for the CNN model development in this study. 
    The orange and blue colors represent \nlikely\ {\likely} and \nunlikely\ {\unlikely} LAE candidates, respectively, as described in Section~\ref{subsec:lc}. 
    The black vertical dashed line shows the Ly\(\alpha\) luminosity of \( \sim 1.0 \times 10^{42.8}~\mathrm{erg~s^{-1}} \)  at \( z \sim 2.7 \), corresponding to  50\% completeness at at a line flux of \( \sim 1.1 \times 10^{-16}~\mathrm{erg~s^{-1}~cm^{-2}} \), which marks the HETDEX survey’s sensitivity limit \citep{Gebhardt2021}.
    }
    \label{fig:distribution_cosmos}
\end{figure*}

Figure~\ref{fig:distribution_cosmos} shows the \lya\ line S/N, 
\lya\ luminosity, and  \lya\ FWHM (measured from a Gaussian model fit to the line emission) distributions for the HETDEX COSMOS catalog and for the labeled subsets described above.

Since the primary goal of this study is to distinguish galaxy-origin emission lines from artifacts, detections of extended \lya\ emission such as Lyman Alpha Nebula (LAN) and identifications of AGN signatures are beyond the scope of this work. 
In the HETDEX survey, LANs are identified using dedicated algorithms described in \citet{Mentuch_Cooper_2026b} and AGNs using those in \citet{Liu2022, Liu2025}.

\begin{deluxetable*}{lccccc}[b]
\tabletypesize{\small}
\tablecaption{Ancillary Dataset Used for Class Label {\likely}\label{tab:Likely_Real}}
\tablehead{
\colhead{Class Label} & & & \colhead{ {\likely}} & &  }
\startdata
Category & Narrow-Band & Spec-$z$ & Photo-$z$ & Repeated Observations & Participatory Science  \\
Datasets & SILVERRUSH,  & COSMOS Spectroscopic & COSMOS2020 & HETDEX & Dark Energy  \\
        & ODIN, and SC4K & Redshift Compilation &            & HDR 5.0.1 & Explorers \\
References & \cite{Kikuta2023}, & \cite{Khostovan2026} & \cite{Weaver2022} & \cite{Gebhardt2021} & \cite{House2023}  \\
 & \cite{Firestone2024}, & & & \cite{Davis2023} &  \\
 &  and \cite{Sobral2018} & & &  & \\
Number counts & 991 & 578 & 911 & 604 & 2841 \\
& (271) & (144) & (349) & (212) & (1691) \\
\hline
Total Count & \multicolumn{5}{c}{\nlikely} \\
& \multicolumn{5}{c}{(2301)} \\
\enddata
\tablecomments{
Numbers in parentheses give the counts in the low-S/N regime ($4.8 \leq \mathrm{S/N} \leq 5.5$). 
The total count reflects the net number of unique sources after accounting for overlaps between the ancillary data sets.}
\end{deluxetable*}
\begin{deluxetable*}{lc}[b]
\tabletypesize{\small}
\tablecaption{Ancillary Dataset Used for Class Label {\unlikely}\label{tab:Unlikely_Real}}
\tablehead{
\colhead{Class Label} & \colhead{ {\it Unlikely Real}} }
\startdata
Category & Participatory Science \\
Datasets & Dark Energy  \\
         & Explorers \\
References & \cite{House2023} \\
Number counts & 4411  \\
& (3941) \\
\hline
Total count & \nunlikely \\
& (3914) \\
\enddata
\tablecomments{
Numbers in parentheses give the counts in the low-S/N regime ($4.8 \leq \mathrm{S/N} \leq 5.5$). The total counts exclude 36 sources (27 in the low-S/N regime) that are identified in multiple ancillary data sets and therefore overlap among the data sets used to define the \likely\ class.}
\end{deluxetable*}
\begin{deluxetable*}{lll}[b]
\tabletypesize{\small}
\tablecaption{Architecture of the CNN Model\label{tab:model_architecture}}
\tablehead{
\colhead{Layer Type} & \colhead{Data Format} & \colhead{Activation}}
\startdata
Input                                                 & (1, 9, 40)  & -- \\
Convolution (kernel=$3\times3$, padding=1)            & (16, 9, 40)  & -- \\
Batch Normalization                                   & (16, 9, 40)  & Leaky ReLU (negative slope=0.1) \\
Dropout (0.2)                                         & (16, 9, 40)  & -- \\
Convolution (kernel=$7\times7$, padding=3)            & (16, 9, 40) & -- \\
Batch Normalization                                   & (16, 9, 40) & Leaky ReLU (negative slope=0.1) \\
Dropout (0.2)                                         & (16, 9, 40) & -- \\
Global Average Pooling                                & (16) & -- \\
Fully connected                                       & (32)        & Leaky ReLU (negative slope=0.1) \\
Fully connected                                       & (1)         & Sigmoid \\
\enddata
\tablecomments{The data format is (channel, width, length).}
\end{deluxetable*}
\begin{deluxetable*}{lrrrrrr}
\tablecaption{Breakdown of the Stratified Dataset\label{tab:stratified}}
\tabletypesize{\scriptsize}
\tablehead{
\colhead{Split} &
\colhead{Number counts} &
\colhead{Fraction} &
\colhead{{\it Likely}, High-S/N} &
\colhead{{\it Likely}, Low-S/N} &
\colhead{{\it Unlikely}, High-S/N} &
\colhead{{\it Unlikely}, Low-S/N}
}
\startdata
Test       & 1697 & 20.0\% & 361 (21.3\%) & 461 (27.2\%) &  92 (5.4\%) & 783 (46.1\%) \\
\cutinhead{Fold 1}
Train      & 4522 & 53.3\% & 962 (21.3\%) & 1226 (27.1\%) & 246 (5.4\%) & 2088 (46.2\%) \\
Validation & 2262 & 26.7\% & 482 (21.3\%) &   614 (27.1\%) & 123 (5.4\%) & 1043 (46.1\%) \\
\cutinhead{Fold 2}
Train      & 4523 & 53.3\% & 963 (21.3\%) & 1227 (27.1\%) & 246 (5.4\%) & 2087 (46.1\%) \\
Validation & 2261 & 26.7\% & 481 (21.3\%) &   613 (27.1\%) & 123 (5.4\%) & 1044 (46.2\%) \\
\cutinhead{Fold 3}
Train      & 4523 & 53.3\% & 963 (21.3\%) & 1227 (27.1\%) & 246 (5.4\%) & 2087 (46.1\%) \\
Validation & 2261 & 26.7\% & 481 (21.3\%) &   613 (27.1\%) & 123 (5.4\%) & 1044 (46.2\%) \\
\enddata
\tablecomments{The training sample (8481 sources) is stratified by class label (\likely/\unlikely) and S/N regime (High: $\mathrm{S/N}>5.5$; Low: $4.8\leq \mathrm{S/N}\leq 5.5$). Percentages in parentheses are computed within each split (i.e., within the Test/Train/Validation set). Owing to rounding, the percentages may not sum to exactly 100\%.}
\end{deluxetable*}

\subsection{Convolutional Neural Network} \label{subsec:cnn}
\subsubsection{Model Architecture} \label{subsec:ma}
CNNs are neural network algorithms that extract spatial features from input data through hierarchical convolutional operations \citep{Cun1989,Krizhevsky2012}. By capturing local features such as edges, textures, and patterns, CNNs are particularly effective in image processing tasks such as segmentation and object recognition \citep[e.g.,][]{Bishop2024, Murphy2022}.
In our case, the 2D spectral images have a simple structure, and the target emission line 
appears the center of each image. This makes CNNs well suited for our task, as their local feature extraction capabilities align with the goal of identifying a single emission line.
While CNNs have limitations in modeling long-range dependencies \citep{Lou2022}, this is not a significant drawback in our setting. 
Models designed to capture global context in large and complex images, such as Transformer-based architectures, are less appropriate for our application.

Transfer learning, which involves applying a pre-developed architecture that is often pre-trained on large-scale image datasets used in computer vision tasks, has recently achieved success in astronomical imaging applications \citep[e.g.,][]{Bhambra2022, Lee2025}.
However, such approaches are not feasible in our case. Pre-developed CNN architectures (e.g., ResNet) typically assume input images of over $100 \times 100$ pixels with three color channels, whereas our 2D spectral images are significantly smaller ($9 \times 40$ pixels) and consist of only a single channel. Consequently, we develop a CNN architecture tailored to our simple 2D spectral input.

Figure~\ref{fig:model_architectures} presents a schematic overview of the  CNN architecture used in this study. 
The model takes a 2D spectral image of size $9 \times 40$ pixels as input and outputs a scalar CNN score between $0.0$ to $1.0$. 
Feature extraction is carried out through two convolutional blocks. The first block consists of a convolutional layer with $16$ channels and a kernel size of $3 \times 3$, followed by batch normalization, a Leaky Rectified Linear Unit (Leaky ReLU) activation, and dropout. The second block has the same structure but uses a larger kernel size of $7 \times 7$.
The resulting feature maps are processed by a global average pooling to produce $16$-channel vectors, which are then passed through a fully connected layer with $32$ hidden units, followed by a Leaky ReLU activation.
The output layer uses a sigmoid activation function to produce a score between 0 and 1, representing the model’s confidence in the presence of an emission line.
This score is not a calibrated probability, as the output is not scaled linearly.
Key architectural parameters, including the number of channels, kernel sizes, batch normalization, and dropout, are summarized in Table~\ref{tab:model_architecture}.
 
We explore several architectural variations to optimize the model's performance. For both the convolutional and fully connected layers, we test different depths, including configurations with 8, 16, and 32 channels. 
Various convolutional kernel sizes are also evaluated, including combinations of $3 \times 3$, $5 \times 5$, $7 \times 7$, and $9 \times 9$, along with different numbers of convolutional blocks, ranging from two to three. 

As for activation functions, we compare ReLU and Leaky ReLU. While ReLU is widely used due to its simplicity and fast convergence, it suppresses all negative input values, which may lead to information loss in noisy data. In contrast, Leaky ReLU allows a small, non-zero gradient in the negative region, enabling the network to preserve subtle negative-valued features that may be astrophysically meaningful and may aid in the identification of \lya\ emission lines.
In our experiments, Leaky ReLU prove to be more robust and sufficient across various configurations, particularly with negative slopes of $0.01$ and $0.1$. 

We also assess the impact of different pooling strategies on feature extraction. 
While max pooling has been successfully employed in many image-classification tasks \citep[e.g.,][]{Ciprijanovic2020, Tadaki2020}, we find that average pooling yields superior performance in our case. This improvement is likely due to the relatively high background noise in the spectral images, where max pooling can occasionally amplify extreme noise, whereas average pooling helps retain diffuse features that are important for identifying \lya\ emission lines \citep[e.g.,][]{Ono2021}. 

The final design, summarized in Table~\ref{tab:model_architecture}, is selected based on its overall stability and classification performance across validation trials. Details of the validation scheme, optimization strategy, loss function, and learning rate scheduling are provided in Section~\ref{subsec:mt}.
Since this study addresses a simple binary classification task using an uncomplicated image-based architecture, more advanced approaches, such as ensembles of classifiers and the Bayesian optimization \citep{Cheng2025, Frazier2018} are left for future work, particularly in the context of fine-grained classification.

\begin{figure*}[t]
\centering
    \includegraphics[trim=0 10 270 20, clip, width=0.8\textwidth]{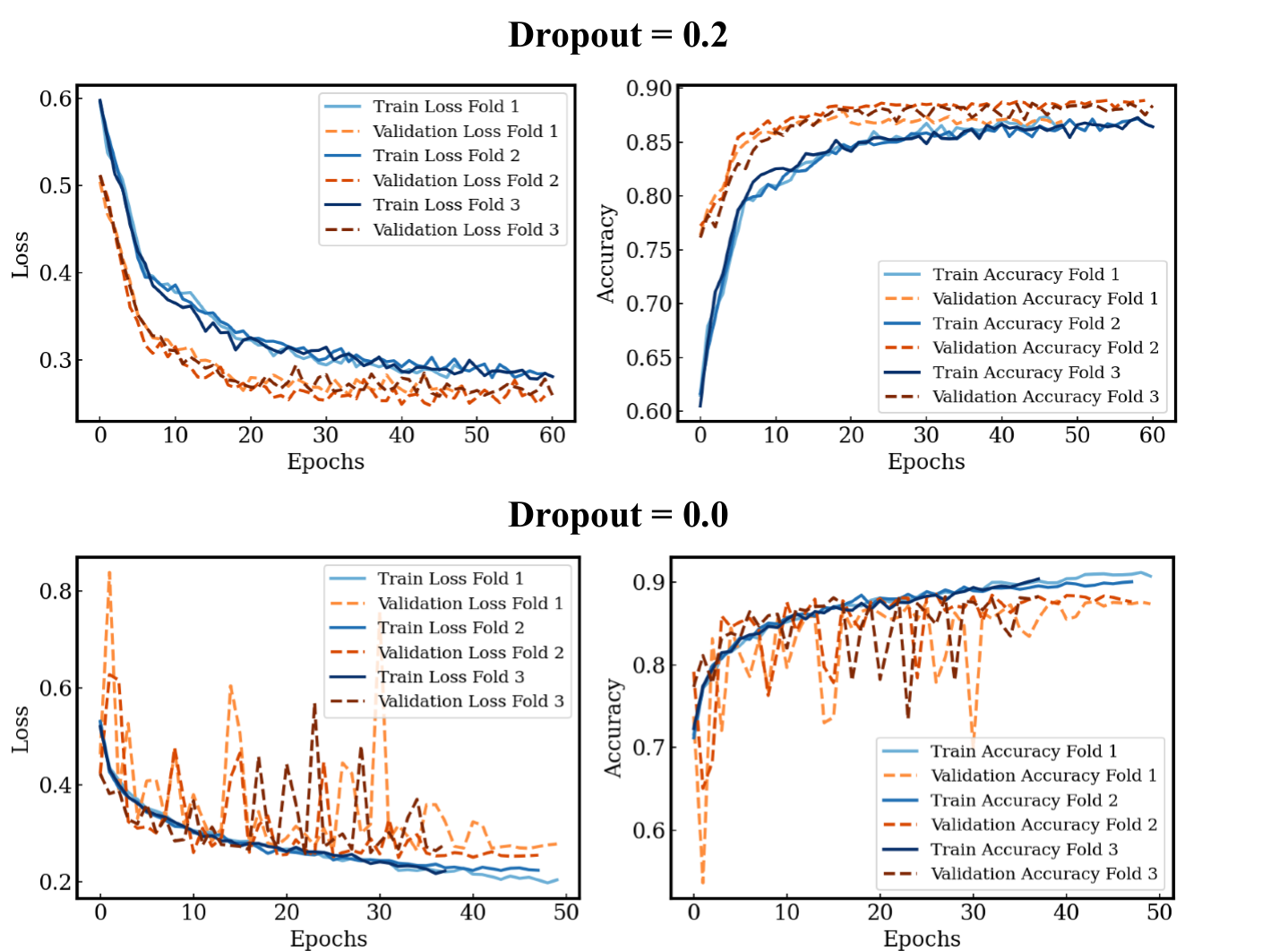}
    \caption{
    {\bf Learning curves of CNN model training with three-fold cross-validation.}
    {\bf Top:} Training and validation loss and accuracy as functions of training epoch for the models trained with dropout. {\bf Bottom:} The corresponding curves for the models trained without dropout. 
    In each panel, the color gradient from dark to light red indicates the validation curves for different folds, while the gradient from dark to light blue represents the corresponding training curves. 
    }
    \label{fig:model_training_curves}
\end{figure*}

\subsubsection{Model Training} \label{subsec:mt}
We train the CNN model using a $k$-fold cross-validation procedure \citep{Kohavi1995,Buchner2024},
which is robust to sample variations in the training data and effectively serves as an ensemble of predictors \citep{Dietterich2000}.
The constructed training sample is first split into $80\%$ for training+validation, and 20$\%$ for testing; for 8481 labeled sources, this corresponds to 6784 and 1697 objects, respectively.
Within each fold, the training+validation subset is further divided into three partitions, with one for validation and the remaining two for training (i.e., three-fold cross-validation).
This process is repeated three times with different validation splits.
Each fold uses $\sim$4523 sources for training and $\sim$2261 for validation.
The final CNN score for each input is computed by averaging the output scores from the three trained models.

For the data split, our constructed training sample exhibits (i) class imbalance between \likely\ and \unlikely\ and (ii) S/N imbalance between the low-S/N ($4.8 \leq \mathrm{S/N} \leq 5.5$) and high-S/N ($\mathrm{S/N} > 5.5$) regimes (Section \ref{subsec:lc}), either of which could bias the model training. We therefore stratify the data by both label and S/N regime to ensure that these distributions are split evenly across the training, validation, and test sets, as well as across the cross-validation folds. The resulting stratified sample sizes are summarized in Table~\ref{tab:stratified}.

Although the data splits are stratified by label and S/N regime across the training, validation, and test subsets, the training subset still has unequal sample sizes across the class--S/N subgroups. As a result, model parameter updates can be dominated by the most populous subgroup.
We therefore mitigate the class imbalance in the training subset using subgroup-dependent augmentation, as follows.
For the low-S/N regime, we equalize the low-S/N counts between classes by leaving the largest subgroup (\unlikely, low S/N) unchanged and augmenting the (\likely, low S/N) subset by adding $\simeq862$ spectral images until it matches the (\unlikely, low S/N) count (2088 2D spectral images in total). 
For the high-S/N regime, we mitigate the class disparity by leaving the (\likely, high S/N) subset unchanged, while augmenting only the smallest subgroup, (\unlikely, high-S/N) by doubling its size (adding 246, for a total of 492 spectral images). For example, in the Fold~1 training split, the subgroup counts change from (962, 1226, 246, 2088) to (962, 2088, 492, 2088) for (\likely, high S/N), (\likely, low S/N), (\unlikely, high S/N), and (\unlikely, low S/N), respectively.
Augmentation is applied only to the training subset, as applying it before the split could introduce data leakage into the validation or test subsets. For each augmented sample, we apply at least one flip, randomly selecting a vertical flip, a horizontal flip, or both with equal probability. 

Overall, the data are stratified by label and S/N regime when split into the training, validation, and test subsets, ensuring comparable distributions across them. In contrast, data augmentation is applied only to the training subset to increase the effective weight of the smaller class--S/N subgroups. 
Although some S/N imbalance remains, our primary focus is on low-S/N sources.
Future work will explore class--S/N-specific training together with more fine-grained classification.

Binary cross-entropy loss is employed for this binary classification task. The Adaptive Moment Estimation (Adam) optimizer \citep{Kingma2015}, which adaptively adjusts learning rates based on moments of the gradients, is employed with an initial learning rate of $0.0001$, 
reduced by a factor of $0.5$ if the validation loss does not improve for five consecutive epochs.  
To enhance generalization and reduce overfitting, L2 regularization \citep{Krogh1991} with a weight decay of $0.0001$ is applied to penalize large weights. Batch normalization \citep{Ioffe2015} and dropout \citep{Srivastava2014} with a dropout rate of $0.2$ are incorporated in the model, yielding the balance between generalization and classification performance. 
The model is trained for up to $150$ epochs, with early stopping triggered if the validation loss does not decrease for $15$ consecutive epochs. The batch size is set to 32.

Figure~\ref{fig:model_training_curves} shows the training and validation loss and accuracy 
as functions of training epoch for each cross-validation fold.
The upper panel corresponds to the model trained with dropout rate $0.2$. In this setting, the validation loss is systematically lower then the training loss and the validation accuracy is higher than the training accuracy. 
This may be attributed to the small validation set and to the use of dropout during training but not during validation. At each epoch, $20\%$ of the units are randomly masked, effectively making training more challenging, whereas validation is performed with dropout disabled and the full network capacity available. The observed curve may suggest that dropout provides effective regularization and may help mitigate overfitting.

The lower panel of Figure~\ref{fig:model_training_curves} shows the corresponding learning curves with dropout disabled (dropout rate $0.0$). In this case, the training and validation loss and accuracy curves become comparable. However, the validation curves exhibit substantial epoch-to-epoch fluctuations, whereas the training curves remain relatively smooth. Without dropout, the model can fit the training set more closely, including both informative patterns and noise, which may increase sensitivity to small parameter updates and lead to unstable validation behavior. This variability may be further amplified by the limited size of the validation set, resulting in larger statistical fluctuations across epochs.

\subsubsection{Performance Metrics} \label{subsec:pm}
Following the model training, we apply the CNN to the test set (1697 2D spectral images) and evaluate its classification performance in both S/N regimes; high S/N ($\mathrm{S/N} > 5.5$) and low S/N ($4.8 \leq \mathrm{S/N} \leq 5.5$). The model performance is assessed using the following metrics:

\begin{itemize}
\item CNN Score Histograms: 
The distribution of the CNN output scores illustrates the model’s ability to separate the two classes.
Histograms are generated for sources labeled as {\likely} and {\it Unlikely Real}, separately for both S/N subsets. 
\item Confusion Matrix:
The confusion matrix \citep{Powers2020} provides the trade-off between true positives and true negatives to evaluate the optimal classification threshold. It is normalized by the total number of sources in each true label category and includes the components TP (true positives), FP (false positives), FN (false negatives), and TN (true negatives). 
\item Receiver Operating Characteristic (ROC) Curve and Area Under the Curve (AUC):
The ROC curve \citep{Fawcett2006} characterizes the trade-off between the true positive rate TP/(TP + FN) and the false positive rate FP/(FP + TN). The AUC serves as a summary measure of classification performance: a value close to $1.0$ indicates strong discriminative power, while $0.5$ corresponds to random guessing.
\item Precision-Recall (PR) Curve and Area Under the Precision-Recall Curve (AUPRC):
For imbalanced data, the ROC curve can appear overly optimistic because the false positive rate could be dominated by the large number of true negatives (even a non-negligible number of false positives can yield a small false positive rate when TN $\gg$ FP) \citep{Saito2015}.
The PR curve \citep{Davis2006} visualizes the trade-off between precision TP/(TP+FP), and recall TP/(TP+FN) (i.e., true positive rate), and is sensitive to performance on the positive class and to contamination among predicted positives.  We therefore compute the PR curve and its summary statistic, the AUPRC, which is informative under class imbalance and complements the ROC curve and AUC.
The optimal classification threshold is calculated by maximizing the F1 score, which is the harmonic mean of precision and recall.
\item Precision Curves:
We also examine how the model’s precision changes with  the CNN score threshold across different S/N limits. This provides a quantitative insight into the model’s performance in both high- and low-S/N regimes.
\end{itemize}

To quantify uncertainties in the performance metrics, 
we fixed the training/validation split within each fold and 
retrained the model using 50 different random seeds. 
This yields a total of 150 trained models (50 seeds $\times$ 3 folds). 
We report 95\% confidence intervals for each metric.

\subsection{Visual Attribution Technique} \label{subsec:vat}
While modern deep learning models achieve high accuracy in computer vision tasks, they often remain non-transparent, with limited interpretability in their decision-making processes.
In this study, we employ a simple and widely used attribution-based visualization method of Grad-CAM++ to visualize the features learned by our CNN models and to gain insight into the model predictions
\footnote{Saliency-based explanation methods are subject to well-known limitations, such as limited sensitivity to model parameters or input data and a lack of quantitative interpretability \citep{Adebayo2018, Rudin2019}. We therefore regard these visualizations as complementary insights rather than definitive explanations. In particular, they are useful for checking whether the model’s focus broadly aligns with human expectations.}.

Gradient-weighted Class Activation Mapping \citep[Grad-CAM;][]{Selvaraju2016}\footnote{For a comprehensive overview and recent advancements in Class Activation Mapping techniques, see, e.g., \citet{Feng2024} and \citet{Gildenblat2021}.}, highlights 
the regions in an input image that are most relevant to a model’s prediction 
by computing the gradient of the predicted class score with respect to the convolutional feature maps.

Grad-CAM++ \citep{Chattopadhyay2017} improves upon Grad-CAM by addressing its coarse object localization and limited ability to deal with multiple object occurrences. It achieves this goal by computing pixel-wise importance weights based on first-, second-, and third-order partial derivatives of the class score with respect to the feature maps: 
\begin{align}
L_{ij}^c &= \sum_{k} w_k^c  A_{ij}^k\label{eq:gradcampp_L}\\
w_k^c &= \sum_{i,j} \alpha_{ij}^{kc} \cdot \text{ReLU}\left( \frac{\partial Y^c}{\partial A_{ij}^k} \right)\label{eq:gradcampp_weight} \\
\alpha_{ij}^{kc} &= 
\frac{
    \frac{\partial^2 Y^c}{(\partial A_{ij}^k)^2}
}{
    2 \cdot \frac{\partial^2 Y^c}{(\partial A_{ij}^k)^2} +
    \sum_{a,b} A_{ab}^k \cdot \frac{\partial^3 Y^c}{(\partial A_{ij}^k)^3}
}\label{eq:gradcampp_alpha},
\end{align}
where $L_{ij}^c$ is the class-specific localization map for class $c$ at spatial location $(i, j)$,
$A_{ij}^k$ denotes the activation at location $(i, j)$ in the $k$-th feature map of convolutional layer, 
$w_k^c$ is its importance weight for class $c$,
$\alpha_{ij}^{kc}$ is a coefficient capturing the contribution of $A_{ij}^k$ to class $c$,
and $Y^c$ is the final score. 
These higher-order weights enable pixel-level attribution within each feature map, resulting in more precise and interpretable visualizations. This approach is particularly effective in cases with multiple contributing regions or diffuse features, such as those observed in astronomical images \citep[][]{Lee2025}.

We apply Grad-CAM++ to each CNN model trained in our three-fold cross-validation and compute the class-specific localization maps for the training set using the weighting scheme defined in Equations~\ref{eq:gradcampp_L}–\ref{eq:gradcampp_alpha}. 
Each Grad-CAM++ map has a shape of $(1, 9, 40)$, identical to the input. 
To examine the relative importance of activations within each image, we normalize each map by rescaling its minimum and maximum values to $0.0$ and $1.0$, respectively.
Since the three folds achieve comparable performance (Figure~\ref{fig:model_training_curves}), the resulting localization maps are consistent across folds. Thus, we present the Grad-CAM++ maps averaged over all folds as a robust representative visualization of the model behavior.

Although Grad-CAM++ is typically applied to the final convolutional layer to highlight class-relevant regions, we also apply it to both the first and second (i.e., final) convolutional layers of our CNN model. Given that our model architecture is simple and straightforward, 
this dual-layer analysis sheds light on the model’s hierarchical operation: 
the first layer captures low-level, edge-like features, while the final layer integrates higher-level semantic information \citep{Zeiler2014}.

By visually inspecting and comparing Grad-CAM++ maps across classification categories (TP, FP, FN, and TN), we qualitatively assess which features the model attends to and whether it consistently focuses on the emission-line regions in the 2D spectral images. 
This provides complementary insights into the model’s behavior rather than definitive explanations.

\begin{figure*}[p]
\centering
    \includegraphics[trim=0 0 0 0, clip, width=1.0\textwidth]{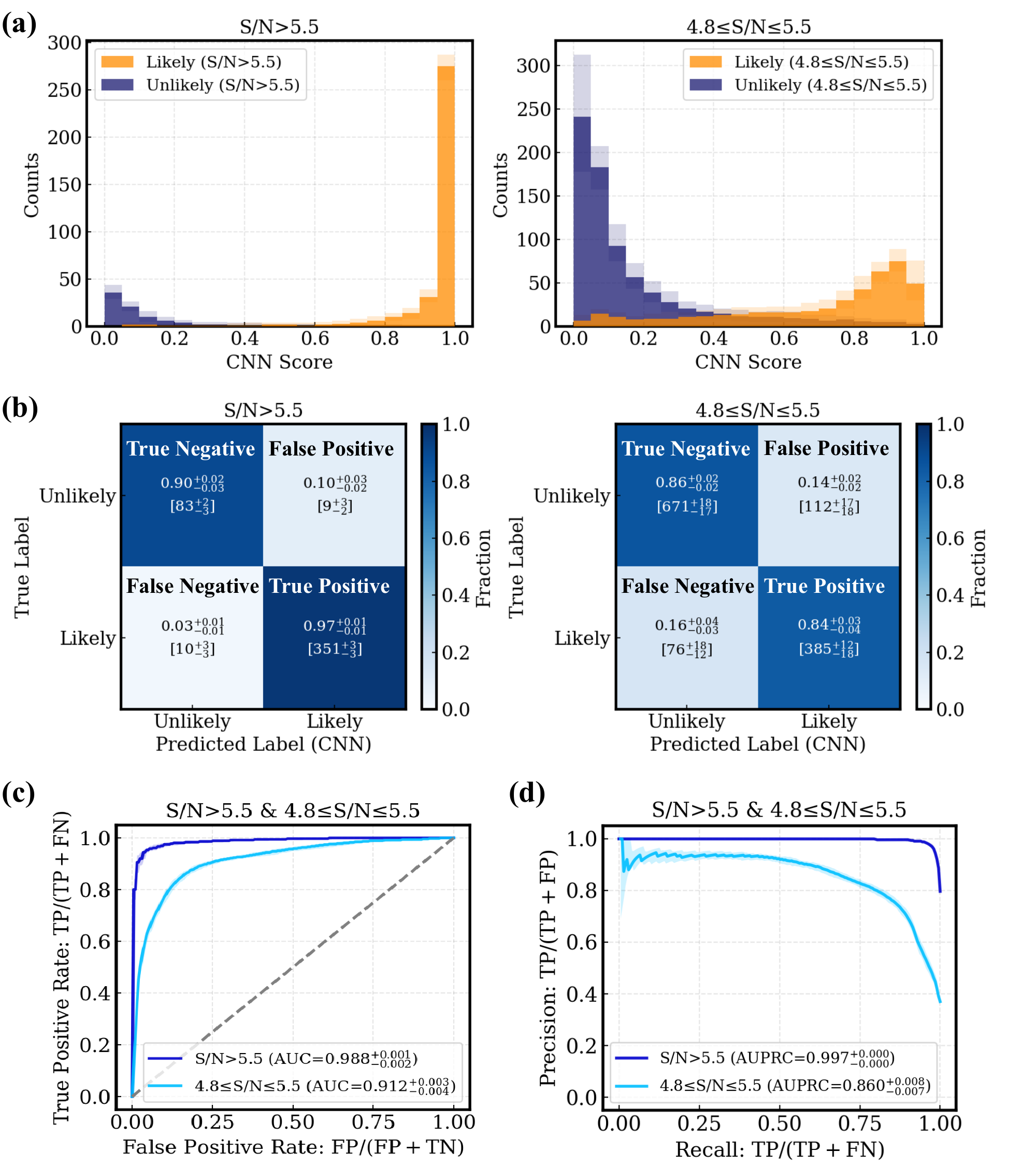}
    \caption{
    {\bf Performance metrics of the CNN model.}
    The trained model is applied to the test set (1697 2D spectral images) drawn from the HETDEX COSMOS catalog.
    {\bf (a)} Distributions of CNN scores. The orange and blue histograms represent sources labeled as {\likely} and {\unlikely}, respectively. The left and right panels show the high-S/N ($\mathrm{S/N} > 5.5$) and low-S/N ($4.8 \leq \mathrm{S/N} \leq 5.5$) subsets.
    {\bf (b)} Confusion matrices at CNN-score thresholds of 0.40 (high-S/N) and 0.37 (low-S/N), presenting the numbers of TP, FP, FN, and TN. The values in each classification (bracket) indicate the fraction normalized by the total number of sources (number counts) in the corresponding true-label category.
    {\bf (c)} ROC curves for the high-S/N (blue) and low-S/N (cyan) subsets; the corresponding AUC values are shown in the inset. The gray dashed line denotes the random-classifier baseline  (AUC $=0.5$).
    {\bf (d)} PR curves for the high-S/N (blue) and low-S/N (cyan) subsets; the corresponding AUPRC values are shown in the inset.
    Shaded regions and error bars indicate 95\% confidence intervals estimated from 150 retrained models (50 different random-seeds realizations of the three-fold procedure).
    }
    \label{fig:model_performance1}
\end{figure*}
\begin{figure*}[p]
\centering
    \includegraphics[trim=-20 0 550 200, clip, width=0.5\textwidth]{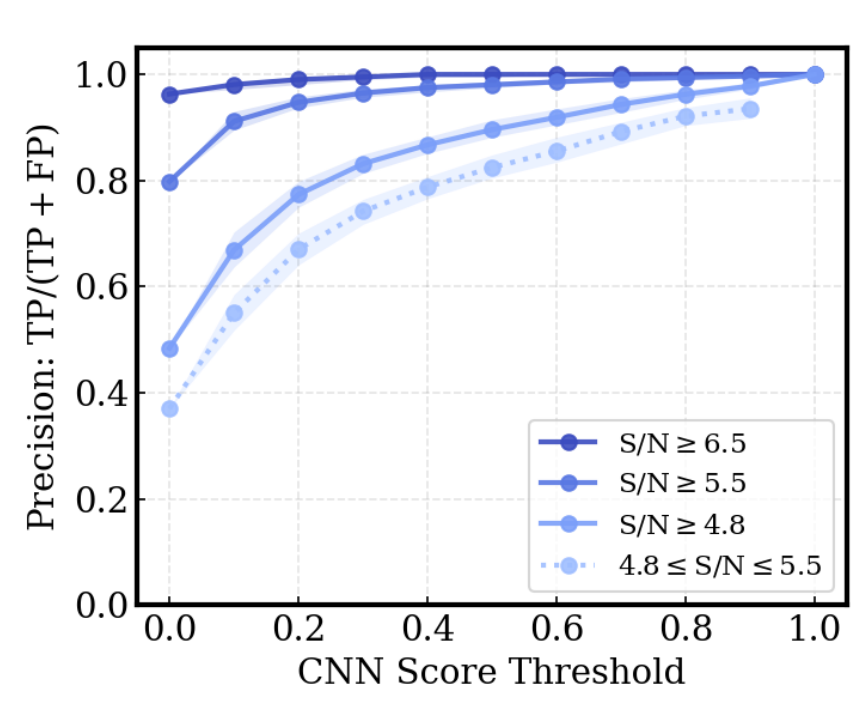}
    \caption{
    {\bf Precision curves as a function of CNN score threshold across various S/N limits.}
    Solid lines indicate subsets defined by S/N lower limits, while dashed line shows an S/N lower limit combined with an upper limit of $\mathrm{S/N} \leq 5.5$. All curves are based on the test set drawn from the HETDEX COSMOS catalog. The shaded regions and error bars indicate 95\% confidence intervals estimated from 150 retrained models (50 different random-seeds realizations of the three-fold procedure).}
    \label{fig:model_performance2}
\end{figure*}
\begin{deluxetable*}{lcccccccc}
\tablecaption{Summary of Classification Performance in Different S/N Regimes}
\label{tab:model_performance}
\tablehead{
\colhead{S/N regime} &
\colhead{F1-optimized threshold} &
\colhead{F1 score} &
\colhead{Precision} &
\colhead{Recall} &
\colhead{Accuracy} &
\colhead{Balanced accuracy} &
\colhead{AUC} &
\colhead{AUPRC} 
}
\startdata
High S/N ($\mathrm{S/N}>5.5$) &
$0.400^{+0.176}_{-0.140}$ &
$0.976^{+0.003}_{-0.003}$ &
$0.975^{+0.011}_{-0.008}$ &
$0.975^{+0.009}_{-0.014}$ &
$0.960^{+0.004}_{-0.004}$ &
$0.941^{+0.018}_{-0.016}$ &
$0.988^{+0.001}_{-0.002}$ &
$0.997^{+0.0004}_{-0.0003}$ \\
Low S/N ($4.8\leq\mathrm{S/N}\leq5.5$) &
$0.372^{+0.100}_{-0.093}$ &
$0.810^{+0.009}_{-0.006}$ &
$0.782^{+0.035}_{-0.043}$ &
$0.844^{+0.032}_{-0.014}$ &
$0.854^{+0.009}_{-0.010}$ &
$0.851^{+0.008}_{-0.006}$ &
$0.912^{+0.003}_{-0.004}$ &
$0.860^{+0.008}_{-0.007}$ \\
Combined ($\mathrm{S/N}\geq4.8$) &
$0.400^{+0.091}_{-0.095}$ &
$0.881^{+0.006}_{-0.005}$ &
$0.870^{+0.024}_{-0.023}$ &
$0.892^{+0.024}_{-0.026}$ &
$0.883^{+0.006}_{-0.007}$ &
$0.883^{+0.006}_{-0.007}$ &
$0.945^{+0.002}_{-0.003}$  &
$0.948^{+0.002}_{-0.003}$  \\
\enddata
\tablecomments{
The classification thresholds are optimized using the F1 score.
Accuracy is defined as $(\mathrm{TP}+\mathrm{TN})/(\mathrm{TP}+\mathrm{TN}+\mathrm{FP}+\mathrm{FN})$,
and balanced accuracy as $(\mathrm{TP}/(\mathrm{TP}+\mathrm{FN})+\mathrm{TN}/(\mathrm{TN}+\mathrm{FP}))/2$.
The uncertainties represent 95\% confidence intervals estimated from 150 retrained models based on 50 random-seed realizations of the three-fold procedure.
}
\end{deluxetable*}
\begin{figure*}[t]
\centering
    \includegraphics[trim=0 0 80 0, clip, width=1.0\textwidth]{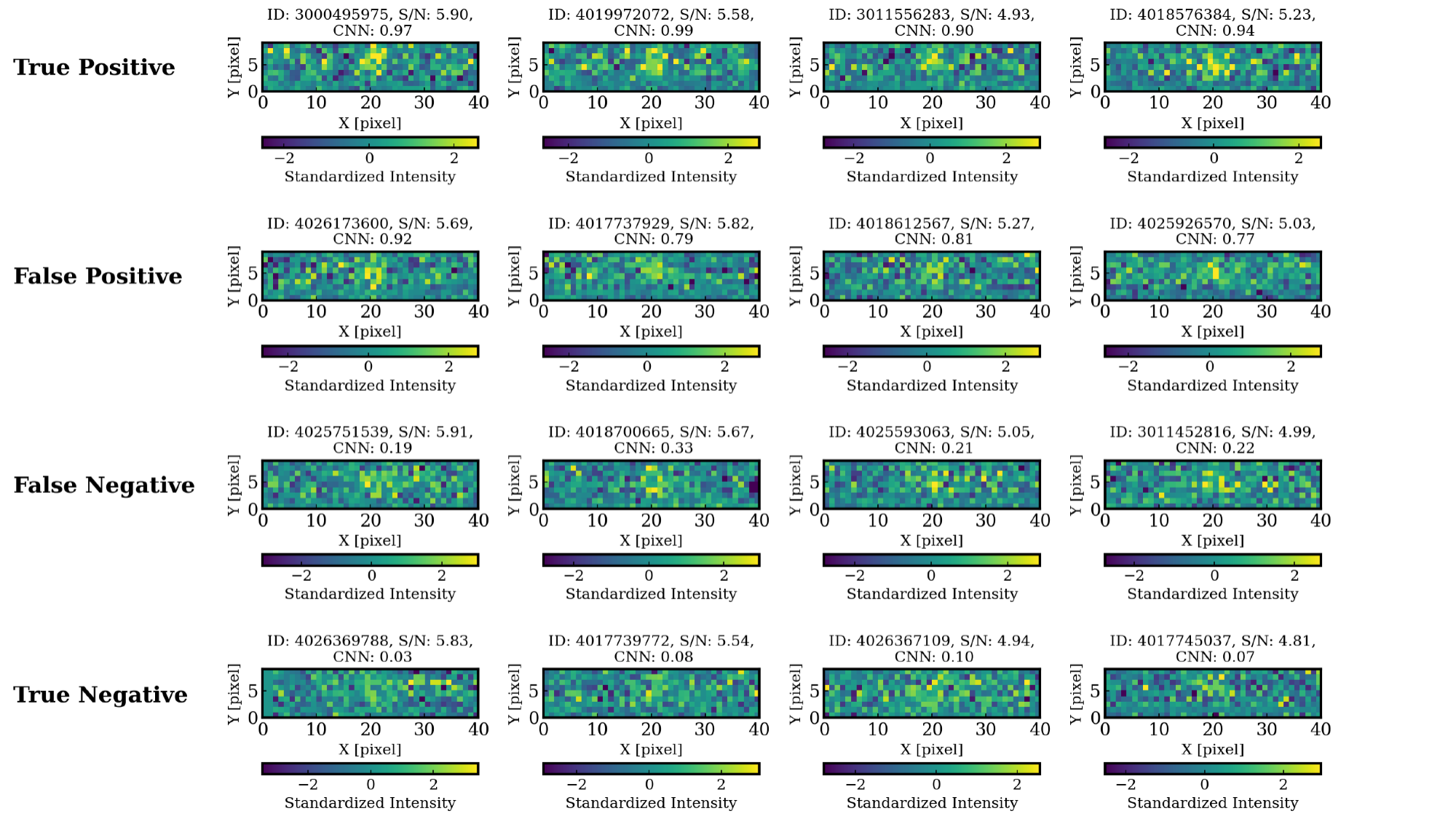}
    \caption{
    {\bf Classification examples for the TP, FP, FN, and TN categories.} 
    Each panel shows representative 2D spectral images randomly selected from each category.
    The HETDEX detection ID and the corresponding S/N and CNN score 
    are displayed at the top of each image. The two left columns show $\mathrm{S/N} > 5.5$, 
    while the two right columns show $4.8 \leq \mathrm{S/N} \leq 5.5$.
    }
    \label{fig:example_2d_spectra}
\end{figure*}

\section{Results and Discussions} \label{sec:results and discussions}
\subsection{CNN Classification} \label{sec:cc}
Figure~\ref{fig:model_performance1} summarizes the performance of the CNN model on the test set in two S/N regimes: high S/N ($\mathrm{S/N} > 5.5$) and low S/N ($4.8 \leq \mathrm{S/N} \leq 5.5$). 
Panel~(a) shows the distributions of CNN scores for the \likely\ and \unlikely\ categories. For individual sources, the standard deviation of the CNN score across 50 realizations of the three-fold procedure with different random seeds is $\lesssim 0.05$. In both the S/N regimes, the score distributions exhibit bimodal-like structures despite the class imbalance.
For comparison within the existing HETDEX classification framework, we present the \elixer\ confidence score distributions for the same test sample in Appendix~\ref{app:elixer}. 
Since \elixer\ is optimized primarily for Ly$\alpha$/non-Ly$\alpha$ classification rather than false-detection rejection, we regard it as a reference metric rather than a baseline.

Panel~(b) presents the confusion matrices. The classification thresholds optimized by F1 scores and the corresponding performance metrics are summarized in Table~\ref{tab:model_performance}. Although the CNN model performs better in the high-S/N regime than in the low-S/N regime, it remains effective in both.
The classification thresholds optimized by F1 scores are 
$0.400^{+0.176}_{-0.140}$ (high S/N) and $0.372^{+0.100}_{-0.093}$ (low S/N), 
yielding F1 scores of $0.976^{+0.003}_{-0.003}$ and $0.810^{+0.009}_{-0.006}$, respectively.
When the two regimes are combined ($\mathrm{S/N} > 4.8$), the F1-optimized threshold is  $0.400^{+0.091}_{-0.095}$ with F1 score $=0.881^{+0.006}_{-0.005}$.
The corresponding precisions are $0.975^{+0.011}_{-0.008}$ (high S/N), $0.782^{+0.035}_{-0.043}$ (low S/N), and $0.870^{+0.024}_{-0.023}$ (combined), while the recalls are $0.975^{+0.009}_{-0.014}$, $0.844^{+0.032}_{-0.014}$, and $0.892^{+0.024}_{-0.026}$, respectively.
The corresponding accuracies, 
$(\mathrm{TP}+\mathrm{TN})/(\mathrm{TP}+\mathrm{TN}+\mathrm{FP}+\mathrm{FN})$, are 
$0.960^{+0.004}_{-0.004}$ (high S/N),
$0.854^{+0.009}_{-0.010}$ (low S/N), and
$0.883^{+0.006}_{-0.007}$ (combined), respectively.
We also compute the balanced accuracies, 
$(\mathrm{TP/(TP+FN)}+\mathrm{TN/(TN+FP)})/2$, which is 
$0.941^{+0.018}_{-0.016}$ (high S/N),
$0.851^{+0.008}_{-0.006}$ (low S/N), and
$0.883^{+0.006}_{-0.007}$ (combined), respectively.

To provide a simple baseline for comparison, a naive classifier that always predicts 
\likely\ would achieve 79.7\% accuracy in the high-S/N test subset (361/92 for \likely/\unlikely; Table~\ref{tab:stratified}). For the low-S/N test subset, a classifier always predicting \unlikely\ would yield 62.9\% accuracy (461/783 for \likely/\unlikely). By contrast, the balanced-accuracy baseline for such a naive classifier is $50\%$. In both S/N regimes, the CNN model outperforms these baselines in accuracy and balanced accuracy, demonstrating that its predictive power is not simply driven by class imbalance.

Panel~(c) presents the ROC curves, with AUC $=0.988^{+0.001}_{-0.002}$ (high S/N) and $0.912^{+0.003}_{-0.004}$ (low S/N). In both regimes, the model performs well above the random-classifier baseline (AUC $=0.5$).
Panel (d) shows the PR curves with AUPRC $=0.997^{+0.0004}_{-0.0003}$ (high S/N) and $0.860^{+0.008}_{-0.007}$ (low S/N). The high AUPRC values indicate that the model maintains relatively high precision over a broad range of recall. At low recall, the precision can change discretely due to the small number of selected samples; a single false positive can produce a visible dip, with a minor effect on the overall AUPRC.

Figure~\ref{fig:model_performance2} shows the precision as a function of the CNN score threshold for various S/N cuts. The precision increases with threshold. For thresholds above 0.5, the model achieves precisions of $0.981^{+0.006}_{-0.006}$ (high S/N), $0.825^{+0.024}_{-0.019}$ (low S/N), and $0.896^{+0.019}_{-0.014}$ (combined). The combined sample reaches a precision of nearly 90\%, suggesting that the use of this threshold can be a practical choice, although the optimal threshold depends on the scientific objective.

Figure~\ref{fig:example_2d_spectra} exhibits classification examples of the TP, FP, FN, and TN categories for the high- and low-S/N regimes.
Higher CNN scores typically correspond to central emission lines with spatially extended structures, while lower scores often reflect sharp or noisy features. To better understand the model’s classification behavior, we examine each category through attribution-based visualization in the following section.

\begin{figure*}[p]
\centering
    \includegraphics[trim=0 5 150 5, clip, width=0.9\textwidth]{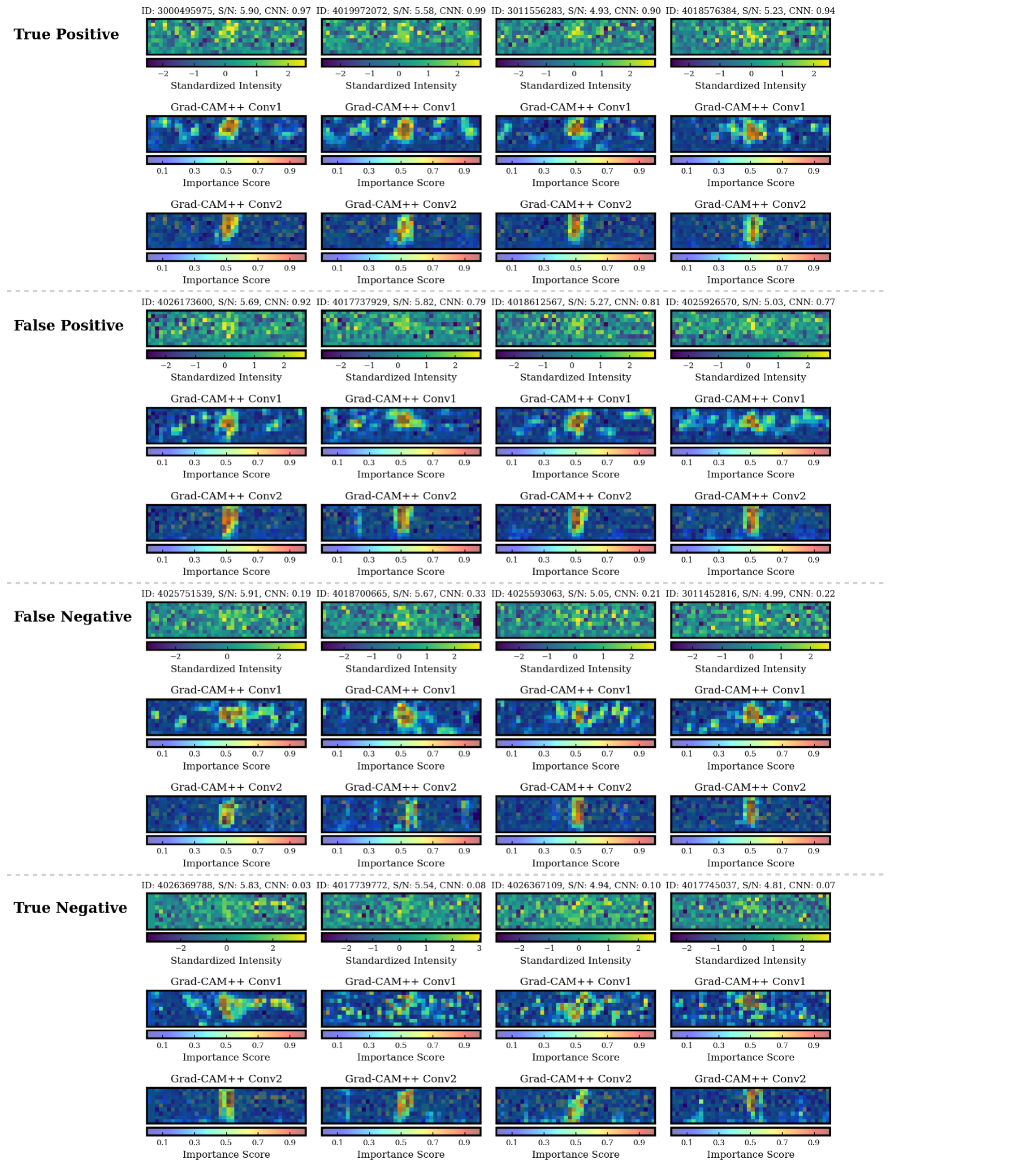}
    \caption{
    {\bf Grad-CAM++ maps for the TP, FP, FN, and TN classifications.}
    Same as Figure~\ref{fig:example_2d_spectra}, but with Grad-CAM++ maps for the convolutional layers of the CNN model used in this study.
    The maps are superimposed on the corresponding 2D spectral images, and the color scale indicates the relative importance of activations within each input image, with values closer to 1.0 representing higher importance.
    }
    \label{fig:example_gradcam_2d}
\end{figure*}
\begin{figure*}[p]
\centering
    \includegraphics[trim=0 0 220 140, clip, width=1.0\textwidth]{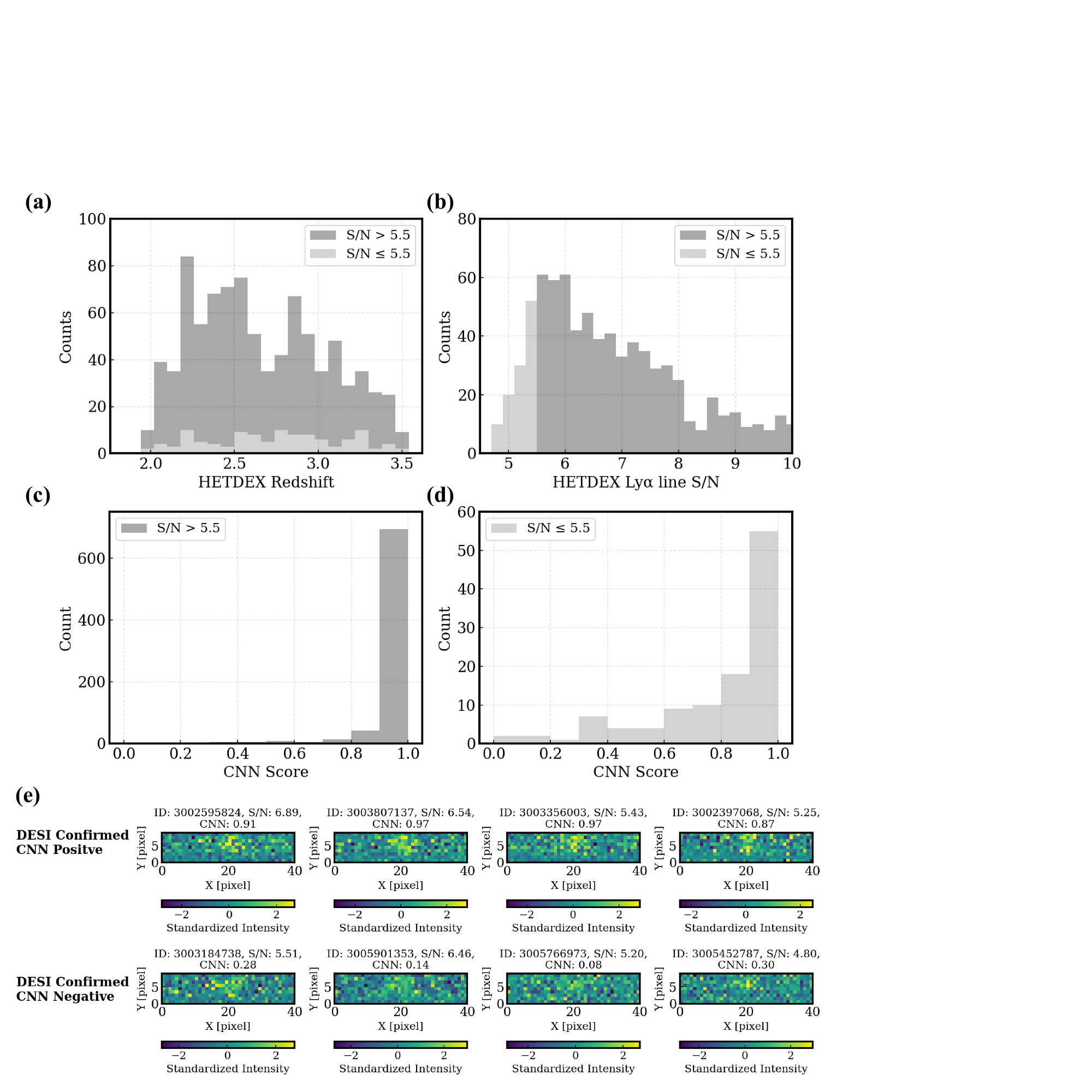}
    \caption{
    {\bf Redshift distribution of DESI-confirmed HETDEX LAEs used in this study, along with histograms of their S/N and CNN scores.}
    The dark and light gray colors indicate the DESI-confirmed HETDEX LAEs with 
    $\mathrm{S/N} > 5.5$ and S/N$\leq5.5$ respectively.
    Panels {\bf (a)} and {\bf (b)} show the HETDEX redshift and S/N distributions, while panels {\bf (c)} and {\bf (d)} present CNN score histograms for $\mathrm{S/N} > 5.5$ and S/N $\leq 5.5$, respectively.
    Panel {\bf (e)} displays examples of 2D spectral images of DESI-confirmed HETDEX LAEs that are classified as positive and negative by the CNN model using the F1-optimized thresholds of 0.40 for the high-S/N, and 0.37 for the low-S/N sources (Section \ref{sec:cc}). These are presented in the same manner as Figure~\ref{fig:example_2d_spectra}.
    }
    \label{fig:distribution_desi}
\end{figure*}

\subsection{Grad-CAM++ Visualization} \label{sec:gv}
Figure~\ref{fig:example_gradcam_2d} presents Grad-CAM++ map examples for the same 2D spectral images shown in Figure~\ref{fig:example_2d_spectra}.
For each classification category (TP, FP, FN, and TN), the top panels display the original 2D spectral images, while the middle and bottom panels show the corresponding Grad-CAM++ maps for the first and second convolutional layers.
The color scale indicates the relative importance of activations within each input image, 
where values closer to 1.0 signify a higher importance attribution.

The Grad-CAM++ maps from the first convolutional layer (Conv1) show two common features across all categories and S/N regimes: activation around the central emission feature and responses distributed along the wavelength direction of the spectral image.
In addition, each category exhibits the following characteristic patterns:
\begin{itemize}
\item \textbf{TP}: Central emission lines with smooth, spatially and spectrally extended profiles, 
extending about 6 pixels along the spectral axis ($\sim$12\,\text{\AA} in the observed frame).  
\item \textbf{FP}: Central features resembling emission lines but with relatively irregular or elongated shapes, accompanied by scattered noise, covering roughly 6--8 pixels along the spectral axis ($\sim$12--16\,\text{\AA} in the observed frame). 
\item \textbf{FN}: Elongated or sharp central emission lines with noise distributed across the spectrum, 
extending over 10--15 pixels ($\sim$20--30\,\text{\AA} in the observed frame). 
\item \textbf{TN}: Irregular or absent central and smooth emission-line features, with noise dominating the spectral images. 
\end{itemize}
These patterns are consistent with the trends identified through visual inspection of the 2D spectral images in Section~\ref{sec:cc}.

The Grad-CAM++ maps from the second convolutional layer (Conv2) are more concentrated on the central part of emission line than those from Conv1. The highlighted region typically indicates a single vertical feature spanning about four pixels along the spectral axis, corresponding to $\sim 8\,\text{\AA}$ in the observed frame, or $\Delta v \approx 430~\mathrm{km~s^{-1}}$ at $z=2.7$. 
To gain further intuition into these patterns from both Conv1 and Conv2, we also present the corresponding pseudo one-dimensional (1D) spectra computed by collapsing the 2D spectral images along the cross-dispersion axis using the mean, as provided in Figure~\ref{fig:example_gradcam_1d} in APPENDIX \ref{app:gradcam_1d}.

The qualitative analyses of the Conv1 and Conv2 maps suggest that positive and negative predictions could be discerned through different uses of spectral information across layers. The Conv1 maps imply that the model could inspect the full wavelength range of the 2D spectra while identifying a spatially and spectrally extended central emission line feature. In contrast, the Conv2 maps are more concentrated around the central feature, suggesting that this layer could serve to further assess the presence and profile of the emission line. These patterns indicate that the model decisions could be guided by physically meaningful features in the input 2D spectral images.

While Grad-CAM++ provides intuitive and spatially informative visualizations, it is inherently limited to qualitative assessments. Future work will incorporate alternative attribution-based techniques, such as SHAP \citep[SHapley Additive exPlanations;][]{Lundberg2017, Alfonzo2024}, to enable quantitative interpretation, particularly for fine-grained classification of the emission-line features.

\subsection{Application to HETDEX DEX Catalog} \label{sec:ath}
We apply our CNN model to 2D spectral images from the HETDEX DEX catalog
to distinguish the LAE candidates from artifacts and sky residuals. 
Section~\ref{subsec:desi} evaluates the recovery of HETDEX LAEs in survey fields beyond COSMOS.
Section~\ref{subsec:rd} examines the redshift distribution of HETDEX LAE candidates in the DEX catalog.
Section~\ref{subsec:cn} presents the cumulative number of LAE candidates obtained by extending the identification into the low-S/N regime, and Section~\ref{subsec:cl} discusses the limitations and potential improvements of our deep learning--based approach.

\subsubsection{Model Performance Beyond the COSMOS Field} \label{subsec:desi}
We investigate the capability of our CNN model to identify emission-line features in survey fields beyond COSMOS, where the model is originally trained. 
Recently, DESI optical spectroscopy by \citet{Landriau2025} confirmed redshifts for approximately 1000 HETDEX LAEs in the HETDEX DEX-Spring field through visual inspection of their \lya\ emission lines. We use these DESI-confirmed LAEs (hereafter DESI-HETDEX LAEs) in our analysis. 

By cross-matching the DESI-HETDEX LAEs with HDR5 data and applying a selection cut of 
a high-confidence flag \texttt{VI\_QUALITY $\geq 3$} and a data quality flag \texttt{DEX\_FLAG $= 1$} provided by \citet{Landriau2025}, we obtain a total of 891 DESI-HETDEX LAEs, consisting of 778 sources with high S/N ($\mathrm{S/N} > 5.5$) and 112 with $4.8 \leq \mathrm{S/N} \leq 5.5$ in the HETDEX DEX-Spring field\footnote{
The DESI-HETDEX LAEs have been originally identified as LAE candidates in the earlier HETDEX Internal Data Release HDR3. Some of these objects are later excluded from HDR5 after application of the latest data quality control.}.

As noted by \citet{Landriau2025}, the overall recovery rate of this dataset cannot be precisely quantified due to inherent observational limitations. 
A key limitation arises from the $1\farcs5$ diameter of the HETDEX fibers and variations in fiber-to-fiber throughput \citep{Gebhardt2021}, 
which can displace the computed centroid of a low-S/N source by more than $1''$ from its true sky position. 
Astrometric uncertainties become increasingly severe at lower S/N ($\mathrm{S/N} \leq 5.5$). 
At the current HETDEX detection limit, the median positional accuracy is $0\farcs6$, with more than $20\%$ of sources offset by over $1\farcs0$. 
Because the DESI spectrograph employs single $1\farcs5$ fibers, any misalignment between the fiber positions and the LAE locations can result in a partial or even complete loss of signal.
Consequently, such sources would not be captured by DESI spectroscopy and are absent from their analysis. 
Thus, this analysis can only confirm that a source is real; it is more difficult to prove that a source is false. Accordingly, the confirmation rate represents a lower limit on the true fraction of real sources. Future work using unbiased IFU observations, such as VLT/MUSE or Keck/KCWI, will be needed to quantify the unconfirmed fraction.

We use the 2D spectral images of the DESI-HETDEX LAEs to compute the distribution of their CNN scores. 
Figure~\ref{fig:distribution_desi} shows the redshift distribution (Panel (a)), as well as histograms of S/N (Panel (b)) and CNN scores (Panels (c) and (d)).
The dark and light gray colors indicate the DESI-HETDEX LAEs with high S/N ($\mathrm{S/N} > 5.5$) and low S/N ($4.8 \leq \mathrm{S/N} \leq 5.5$), respectively.
Adopting the F1-optimized thresholds from Section \ref{sec:cc},
our CNN model recovers $99\%$ of DESI-HETDEX LAEs with high S/N and $93\%$ of those with low S/N  at thresholds of 0.40 and 0.37, respectively.
Panel (e) displays representative 2D spectral images of DESI-confirmd LAEs 
classified as positive and negative by the CNN model using the thresholds.
A small fraction (a few percent) of spectral images receive low CNN scores. These cases exhibit sharp and noisy central emission-line features consistent with the signatures of FN predictions seen in the Grad-CAM++ maps in Section~\ref{sec:gv}. These noisy structures may partly reflect data taken under poor observing conditions during some HETDEX observations.

Overall, these results indicate that our CNN model, trained on the COSMOS field, provides consistent score-assignment performance for LAE candidates in the independent survey field.

\begin{figure}[t]
\centering
    \includegraphics[trim=0 0 500 200, clip, width=0.5\textwidth]{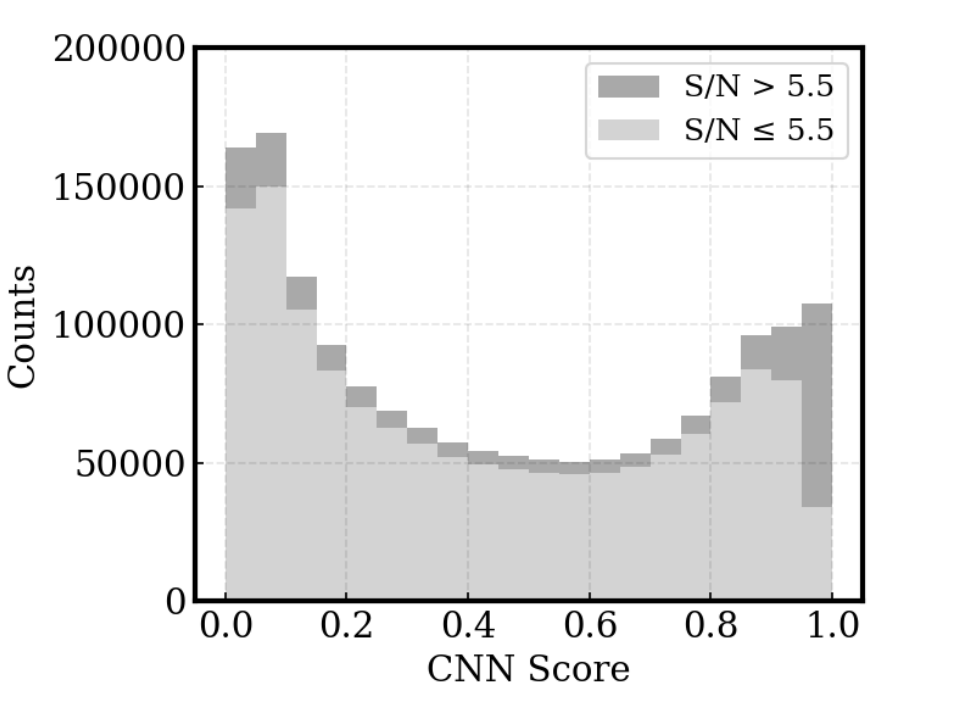}
    \caption{
    {\bf Distribution of CNN scores for LAE candidates in the HETDEX DEX catalog across different S/N ranges.} 
    The dark and light gray indicate sources with $\mathrm{S/N} > 5.5$ and $\mathrm{S/N} \leq 5.5$, respectively. 
    }
    \label{fig:cnn_score_distribution}
\end{figure}
\begin{figure*}[t]
\centering
    \includegraphics[trim=0 0 0 180, clip, width=0.95\textwidth]{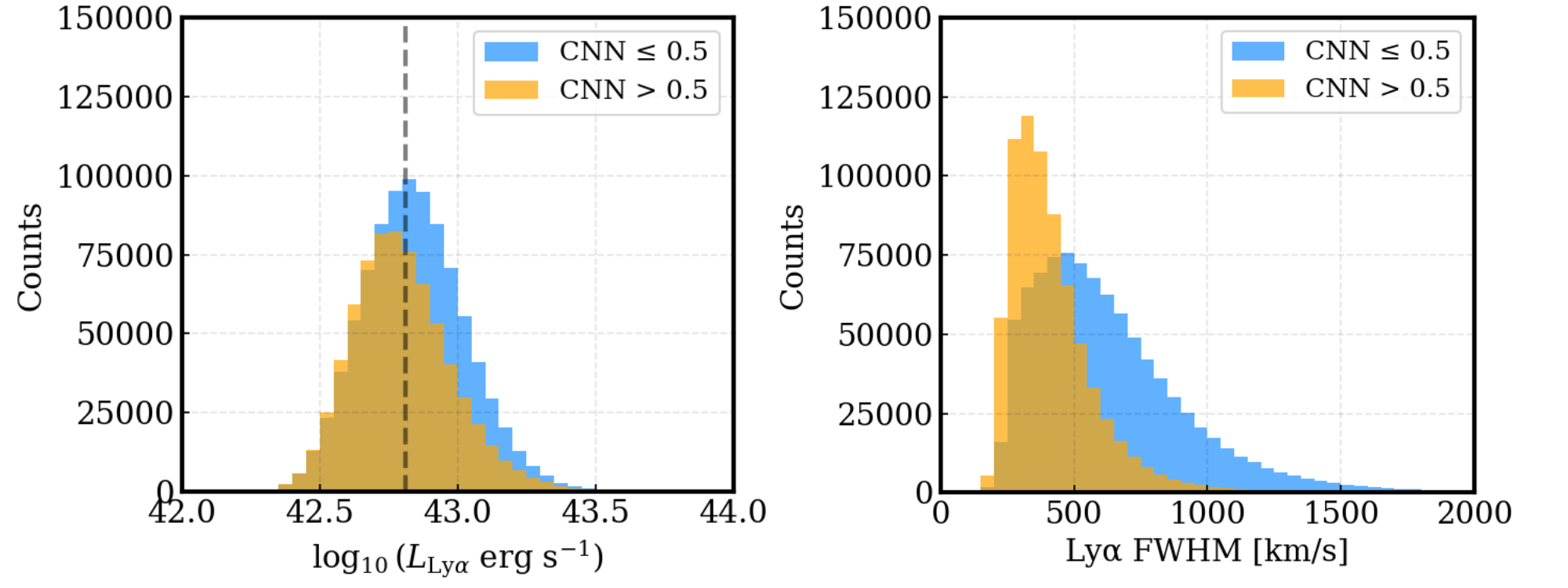}
    \caption{
    {\bf Distribution of \lya\ luminosity and FWHM} for LAE candidates in the HETDEX DEX catalog at a CNN score threshold of 0.5.
    The yellow and blue colors indicate LAE candidates with a CNN scores $> 0.5$ and $\leq 0.5$, respectively.
    The vertical black dashed line in the left panel marks the Ly\(\alpha\) luminosity of 
    $\sim 1.0 \times 10^{42.8}~\mathrm{erg~s^{-1}}$ at $z \sim 2.7$, which corresponds to the HETDEX survey’s sensitivity limit, as in Fig~\ref{fig:distribution_cosmos}.}
    \label{fig:distribution_dex}
\end{figure*}
\begin{figure*}[p]
\centering
    \includegraphics[trim=0 0 250 0, clip, width=0.8\textwidth]{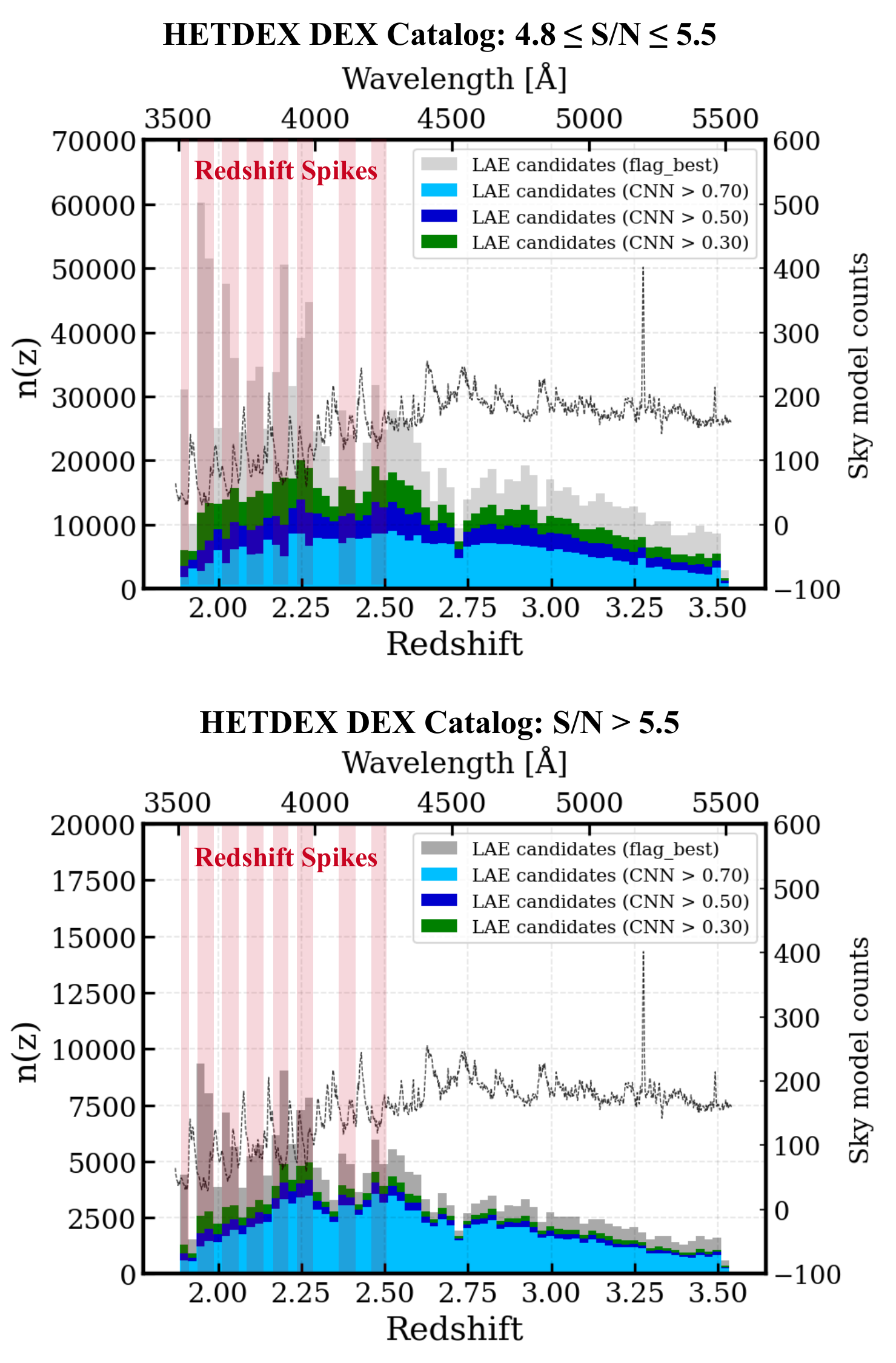}
    \caption{
    {\bf Redshift distributions of LAE candidates selected with different CNN score thresholds in the HETDEX DEX catalog.}
    Histograms showing the redshift distribution of LAE candidates selected by data-quality flags alone (gray; same as the left panel of Figure~\ref{fig:flag_best}), and by applying additional CNN score thresholds of 0.70 (cyan), 0.50 (blue), and 0.30 (green). 
    The top and bottom panels correspond to sources with $4.8 \leq \mathrm{S/N} \leq 5.5$ and $\mathrm{S/N} > 5.5$, respectively. The bin width is $\Delta z = 0.025$. 
    The red-shaded regions highlight spurious redshift spikes, and the dashed lines represent the sky model in the HETDEX survey. Both correspond to those shown in the left panel of Figure~\ref{fig:flag_best}.
    }     
    \label{fig:dndz_dex}
\end{figure*}
\begin{figure}[t]
\centering
    \includegraphics[trim=0 0 180 220, clip, width=1.0\textwidth]{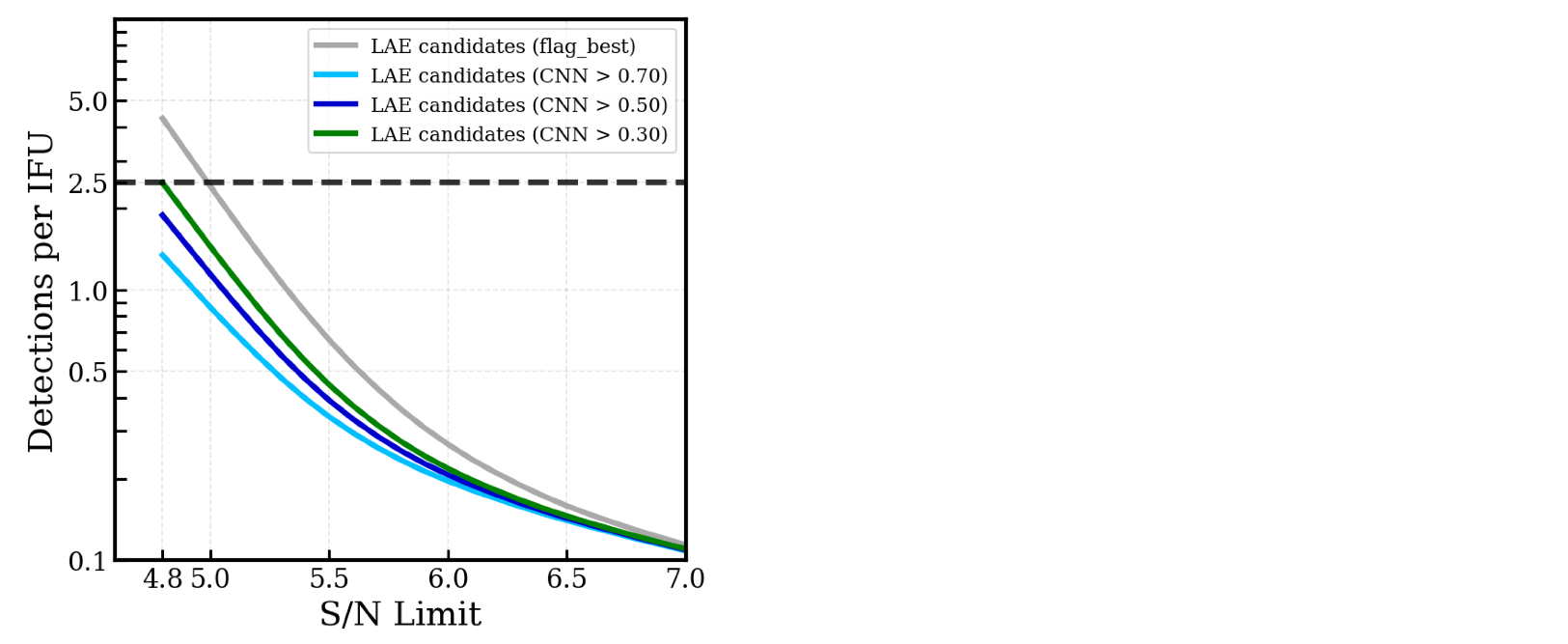}
    \caption{
    {\bf Cumulative number of LAE candidates selected with different CNN score thresholds in the HETDEX DEX catalog.}
    Same as the right panel of Figure~\ref{fig:flag_best}, but for LAE candidates selected using CNN score thresholds of 0.70 (cyan), 0.50 (blue), and 0.30 (green), along with data quality flags (gray). Each curve shows the cumulative number of LAE candidates per VIRUS IFU as a function of the S/N threshold. The dashed line marks the target LAE density adopted in the HETDEX survey \citep{Gebhardt2021}. 
    }
    \label{fig:dn_dex}
\end{figure}

\subsubsection{Redshift distribution of HETDEX LAEs} \label{subsec:rd}
We apply the CNN model to a total of {\ndexuse} 2D spectral images of the LAE candidates from the HETDEX DEX catalog. 
Figure~\ref{fig:cnn_score_distribution} shows the CNN score histograms across different S/N regimes. The distributions exhibit a bimodal-like structure, similar to that observed in the training sample.
Figure~\ref{fig:distribution_dex} presents the distributions of \lya\ luminosity and FWHM for LAE candidates selected at a CNN score threshold of 0.5, where the overall precision reaches nearly $90\%$ when the high-S/N and low-S/N regimes are combined (Section~\ref{sec:cc}).
Both the distributions closely resemble those of the COSMOS LAE candidates (Figure~\ref{fig:distribution_cosmos}). 
The CNN model is trained exclusively on the COSMOS catalog, using a limited training sample ($\sim5000$ sources), which is $\sim1/300$ of the size of the DEX catalog. Such a disparity could limit generalization to DEX. 
However, the similarity between the DEX and COSMOS distributions likely reflects the uniform observing strategy of HETDEX, which ensures relatively homogeneous data quality, although variations in observing conditions and instrumental performance introduce some degree of inhomogeneity \citep{Gebhardt2021}.
Taken together, these results suggest that the CNN model can be applied to LAE candidates in the HETDEX DEX catalog beyond COSMOS.

Figure~\ref{fig:dndz_dex} presents the redshift distributions of LAE candidates selected with different CNN score thresholds in the HETDEX DEX catalog.
The CNN model filters out approximately 
$69\%$ (1,122,221/1,632,398) of sources at a threshold of 0.70,
$56\%$ (916,211/1,632,398) at 0.50, and 
$42\%$ (689,670/1,632,398) at 0.30 
from the DEX catalog over the entire S/N range. Notably, the model also substantially suppresses 
redshift spikes at $z\sim 1.9\ –\ 2.5$, thereby mitigating possible contamination and yielding a smoother LAE distribution across the survey redshift range.

\subsubsection{Cumulative Number of HETDEX LAEs} \label{subsec:cn}
Figure~\ref{fig:dn_dex} shows the cumulative number of LAE candidates per IFU as a function of the S/N limit for the different CNN score thresholds, compared with the data quality–only selection.
The CNN model extends the identification of LAE candidates into $4.8 \leq \mathrm{S/N} \leq 5.5$, while assigning lower scores to candidates with less convincing spectral features across the entire S/N range, including at $\mathrm{S/N} > 5.5$.
The cumulative number of LAE candidates indicates that a CNN score threshold of 0.3 reaches the survey’s target LAE density of 2.5, although this threshold is likely to include a higher level of contamination. 
In contrast, a threshold of 0.5 may provide a practical threshold by reducing contamination while retaining a large LAE candidate sample in the DEX catalog, with a cumulative number density of $\sim2.0$, 
close to the target value although slightly below it.

\subsubsection{ Current Limitations and Potential Improvements} \label{subsec:cl}
The slight shortfall relative to the survey target density seen in the cumulative number of LAE candidates may indicate that the CNN model does not yet fully capture the characteristics of \lya\ emission lines under the noisy conditions of the observed data.
Two primary factors contribute to this limitation: the predictive uncertainty of the model and the limited representativeness of the training set\footnote{A highly sophisticated model that employs Dirichlet head and normalizing flow, exclusively disentangling these uncertainty estimations, will be presented in a forthcoming paper (Shen et al.\ in prep.).}.
The predictive uncertainty of the CNN score is evaluated by a Monte Carlo method (Appendix~\ref{app:std_cnn}), and estimated to be $\lesssim0.17$ for S/N~$\lesssim7$.
With or without this uncertainty, a non-negligible fraction of sources remain ambiguously classified (e.g., CNN scores of $0.3\ –\ 0.7$ in Figure~\ref{fig:cnn_score_distribution}), suggesting that such cases are inherently challenging for the model and may be underrepresented in training.

For instance, the observed \lya\ line profiles can vary due to both redshift-dependent variations and intrinsic physical diversity \citep[][]{Ouchi2020}.
The redshift variations over $z = 2$–$3$ may have little impact on the CNN’s score assignment, 
as the input spectra are centered on the detected emission and the visual attribution analysis indicates that the model primarily focuses on the central part of the emission line.
In contrast, even within similar redshift ranges, the physical diversity of \lya\ emission can be subtle in low resolution spectra and easily obscured by noise, which may lead to the ambiguous predictions. 
Even if these factors are mitigated, developing a perfect classifier that can distinguish true emission lines from false detections based solely on a single emission line in untargeted spectroscopic surveys remains unrealistic, as such a model would likely overfit to specific samples.

Our best practice for utilizing the CNN scores is to adopt a relatively relaxed threshold and combine it with complementary statistical or machine-learning methods to further recover LAE candidates. 
At a threshold of 0.5, the model attains a precision close to $90\%$ across the entire S/N range (Figure \ref{fig:model_performance2}), broadly consistent with the $10\%$ false-detection requirement \citep[][]{Gebhardt2021}
\footnote{
This requirement is distinct from the separate HETDEX science requirement that contamination of the LAE sample by low-redshift interlopers, primarily [O~II] emitters, remain below 2\% \citep{Gebhardt2021}. The present CNN analysis is not intended to address that LAE/[O~II] classification requirement.
The interloper-contamination requirement is largely satisfied in the \elixer\ framework, for which \citet{Davis2023} reported a projected [O~II] contamination rate of 1.2\% in the LAE sample.}, 
although the optimal threshold should ultimately be chosen by balancing the gain in LAE sample size against the increase in false-positive contamination. Potentially effective complementary approaches include statistical removal of systematic residuals, random forest classifiers based on emission-line source properties (MC26), and unsupervised clustering of the HETDEX spectra \citep[][]{House2024}. 

One promising direction is the use of self-supervised learning \citep[][]{Huertas-Company2023b, Parker2024, Desmons2024}. Contrastive learning, for instance, embeds large data volumes into a meaningful latent space by drawing similar examples closer and pushing dissimilar ones apart. Such pre-training method has been shown to outperform supervised learning in downstream tasks (e.g., classification), even when only a small fraction of labeled data is available \citep{Siudek2025, Hayat2021, Stein2021}.

\section{Conclusions} \label{sec:conclusions}
In this work, we present a deep learning approach for enhancing LAE identification and suppressing false detections in the HETDEX untargeted spectroscopic survey, which obtains optical spectra over wide sky areas without imaging pre-selection. We develop a  CNN model that operates on 2D spectral images of \ndexuse\ \lya\ emission-line candidates at $1.9 < z < 3.5$ across the $87~\mathrm{deg}^2$ survey area. Our model addresses the following key challenges:

\begin{enumerate}[label=(\roman*)]
    \item Low-S/N classification: To distinguish the \lya\ emission-line candidates from artifacts and sky residuals primarily in the low-S/N regime ($4.8 \leq \mathrm{S/N} \leq 5.5$), we use 2D spectral images that encode both spectral (dispersion) and cross-dispersion profiles on the detector plane.
    \item Data scale and resolution: The model is 
    designed for small-format (1 channel $\times$ 9~$\times$~40 pixels), low-resolution ($R\sim800$) 2D spectral images, for which transfer-learning approaches are not well suited.
    \item Limited label availability: We construct a fiducial custom-labeled training sample utilizing ancillary observational data in the COSMOS field and leveraging diagnostics contributed by our participatory science project \textit{Dark Energy Explorers}.    
    \item Interpretability: We employ a simple and widely used visual attribution technique, Grad-CAM++, to highlight the spectral–spatial features attended to by the model.    
\end{enumerate}

The CNN achieves a balanced accuracy, precision, and recall of $94.1\%$, $97.5\%$, and $97.5\%$ in the high-S/N regime ($\mathrm{S/N}>5.5$), and $85.1\%$, $78.2\%$, and $84.4\%$ in the low-S/N regime. For the combined sample with $\mathrm{S/N}\geq4.8$, the corresponding values are $88.3\%$, $87.0\%$, and $89.2\%$, respectively. 
Using HETDEX LAEs independently confirmed by DESI spectroscopy, the model recovers $99\%$ and $93\%$ of the high- and low-S/N LAEs, respectively, providing an external assessment of model performance. 
Grad-CAM++ attribution maps show that true positives exhibit smooth, spectrally and spatially extended central emission, while true negatives show irregular or absent emission-line features accompanied by noise across the spectra.

Applied to 2D spectral images from the full HETDEX catalog, the CNN extends confident LAE identification down to the low-S/N regime while suppressing spurious redshift spikes over $z\sim 1.9$--$2.5$, where instrumental sensitivity and noise modeling affect the redshift distribution. 
At a practical threshold of 0.5, which yields a precision of nearly $90\%$ in the test sample, the model retains a large fraction of LAE candidates in the catalog while keeping the cumulative number density close to the survey target. The CNN application paves the way for HETDEX cosmological analyses by increasing the effective sample size while mitigating false positives in galaxy correlation measurements.

Our work demonstrates the value of domain-specific deep learning for low-S/N spectroscopic identification and its potential to enhance the scientific yield of large-scale, untargeted spectroscopic surveys.

\section*{Acknowledgements}

We thank the anonymous referee for constructive comments and suggestions that helped to improve the manuscript. We are grateful to Meredith C. Powell, Hasti Khoraminezhad, Donghui Jeong, Eiichiro Komatsu, and Delaney A. Dunne for insightful input from cosmological analyses; and Satoshi Kikuta for making the LAE catalog available.

HETDEX is led by the University of Texas at Austin McDonald Observatory and Department of Astronomy with participation from the Ludwig-Maximilians-Universit\"at M\"unchen, Max-Planck-Institut f\"ur Extraterrestrische Physik (MPE), Leibniz-Institut f\"ur Astrophysik Potsdam (AIP), Texas A\&M University, The Pennsylvania State University, Institut f\"ur Astrophysik G\"ottingen, The University of Oxford, Max-Planck-Institut f\"ur Astrophysik (MPA), The University of Tokyo, and Missouri University of Science and Technology. In addition to Institutional support, HETDEX is funded by the National Science Foundation (grant AST-0926815), the State of Texas, the US Air Force (AFRL FA9451-04-2-0355), and generous support from private individuals and foundations.

Observations were obtained with the Hobby-Eberly Telescope (HET), which is a joint project of the University of Texas at Austin, the Pennsylvania State University, Ludwig-Maximilians-Universit\"at M\"unchen, and Georg-August-Universit\"at G\"ottingen. The HET is named in honor of its principal benefactors, William P. Hobby and Robert E. Eberly.

VIRUS is a joint project of the University of Texas at Austin, Leibniz-Institut f\"ur Astrophysik Potsdam (AIP), Texas A\&M University (TAMU), Max-Planck-Institut f\"ur Extraterrestrische Physik (MPE), Ludwig-Maximilians-Universit\"at Muenchen, Pennsylvania State University, Institut fur Astrophysik G\"ottingen, University of Oxford, and the Max-Planck-Institut f\"ur Astrophysik (MPA). In addition to Institutional support, VIRUS was partially funded by the National Science Foundation, the State of Texas, and generous support from private individuals and foundations.

The authors acknowledge the Texas Advanced Computing Center (TACC) at The University of Texas at Austin for providing high performance computing, visualization, and storage resources that have contributed to the research results reported within this paper. URL: http://www.tacc.utexas.edu

Dark Energy Explorers is recognized as an official NASA Citizen Science partner. This publication utilizes data generated through the Zooniverse.org platform, the development of which is supported by generous funding, including a Global Impact and Award from Google a grant from the Alfred P. Sloan Foundation.

The results of the Dark Energy Explorers would not be as robust and useful if not for the care and dedication of the volunteers. We are extremely grateful for their work. It is having a large impact and is motivational.

The Institute for Gravitation and the Cosmos is supported by the Eberly College of Science and the Office of the Senior Vice President for Research at the Pennsylvania State University.

K.G. acknowledges support from NSF-2008793. 
S.S. acknowledges support from the National Science Foundation under grants NSF-2219212 and NSF-2511145. 
C.G. acknowledges support from the National Science Foundation under grant AST-2408358.

\facility{The Hobby-Eberly Telescope (McDonald Observatory)}
\software{This research was made possible by the open-source projects 
\texttt{hetdex-api} (\url{https://github.com/HETDEX/hetdex_api}), \texttt{astropy} \citep{astropy:2018}, 
\texttt{pytorch} \citep{Paszke2019}, 
\texttt{scipy} \citep{scipy}, \texttt{numpy} \citep{harris2020array}, 
and \texttt{python} \citep{pythonref}.}

\appendix
\counterwithin{figure}{section}

\section{Comparison with ELiXer $p(\mathrm{Ly}\alpha)$\ Score}\label{app:elixer}
As a comparison with the existing HETDEX classification framework, we compare our CNN score with the \elixer\ $p(\mathrm{Ly}\alpha)$ score for the same test sample. The $p(\mathrm{Ly}\alpha)$ score, listed in the HETDEX catalogs as \texttt{plya\_classification}, is a confidence score, not a proper probability, ranging from 0 to 1, with higher values indicating stronger support that the detected emission line is Ly$\alpha$ \citep{Davis2023, Melchior2023}. 

We clarify, however, that \elixer\ is not a direct baseline because the two quantities are optimized for related but not identical tasks. The $p(\mathrm{Ly}\alpha)$ score is primary developed to distinguish Ly$\alpha$ emitters from non-Ly$\alpha$ emission-line sources, in particular low-redshift [O~II] interlopers. In addition, the \elixer\ also incorporates disqualification logic for other emission lines and for problematic detections (including meteors, reduction problems, and poor observations) so that such cases receive lower $p(\mathrm{Ly}\alpha)$ values or be effectively disfavored within the classification framework. By contrast, our CNN is designed to operate one step earlier in the workflow. Rather than performing line-identity classification directly, it is optimized to reject noise/artifact-driven false detections among \elixer-selected Ly$\alpha$ candidates. Therefore, \elixer\ is not treated here as a baseline model for our specific task.

Figure~\ref{app:elixer} shows the score distributions of the \elixer\ $p(\mathrm{Ly}\alpha)$ for the same test data as that used in panel~(a) of Figure~\ref{fig:model_performance1}. 
We focus on the score histograms, rather than confusion matrices or ROC/PR curves, because the  $p(\mathrm{Ly}\alpha)$ values for the HETDEX Ly$\alpha$ candidates are concentrated near the high-score end. In this regime, histogram-based comparisons provide a clearer view of how the two methods behave on the same set of detection candidates.
Compared with the $p(\mathrm{Ly}\alpha)$ score, the CNN score spans a broader dynamic range and exhibits stronger separation between \likely\ and \unlikely\ classes, consistent with its intended role as a false-detection rejection step in the HETDEX Ly$\alpha$-candidate identification workflow.

\begin{figure*}[t]
\centering
    \includegraphics[trim=0 0 0 180, clip, width=0.95\textwidth]{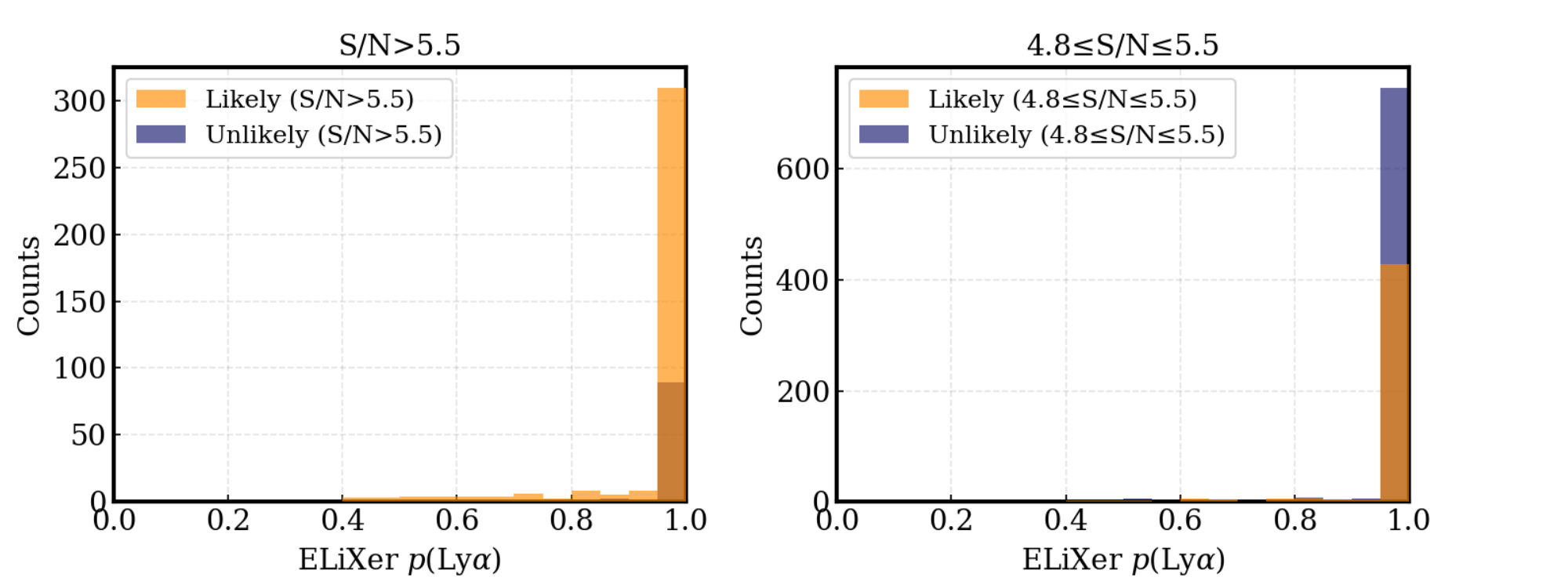}
    \caption{
    {\bf Score Distributions of the \elixer\ $p(\mathrm{Ly}\alpha)$.}
    The plot format is the same as that of panel~(a) in Figure~\ref{fig:model_performance1}.     
    The $p(\mathrm{Ly}\alpha)$ values for LAE candidates in both the \likely\ and \unlikely\ classes are concentrated toward the high-score end.}  
    \label{app:elixer}
\end{figure*}

\section{Collapsed 1D Spectra for Grad-CAM++ Maps}\label{app:gradcam_1d}
\begin{figure*}[b]
\centering
    \includegraphics[trim=0 0 200 0, clip, width=0.9\textwidth]{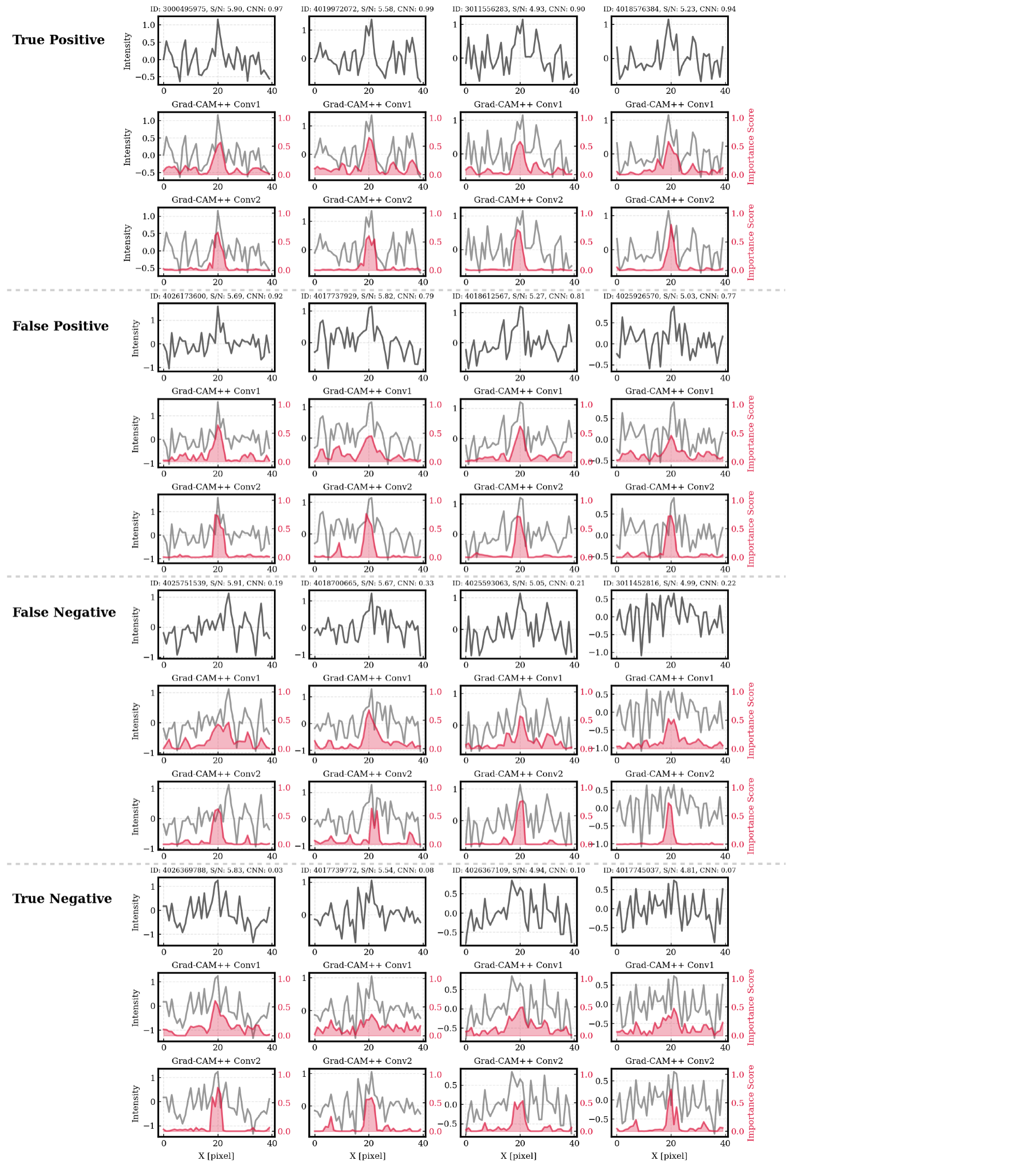}
    \caption{
    {\bf Pseudo 1D Grad-CAM++ maps for the TP, FP, FN, and TN classifications.}
    Same as Figure~\ref{fig:example_gradcam_2d}, but for the pseudo–1D spectra derived from the 2D spectral images and their Grad-CAM++ maps. Each spectral pixel corresponds to 2${\rm \AA}$. The Grad-CAM++ maps (red) are overlaid on the corresponding pseudo–1D spectra (gray) using the same panel layout. Values closer to 1.0 indicate higher relative activation importance. Shown for qualitative purposes only.
    }
    \label{fig:example_gradcam_1d}
\end{figure*}

Figure~\ref{fig:example_gradcam_1d} shows the pseudo–1D spectra corresponding to 
Figure~\ref{fig:example_gradcam_2d}, obtained by collapsing the 2D spectral images along the 
spatial (cross-dispersion) axis using a mean operation. 
These spectra are provided for visual inspection only. The model’s predictions are primarily guided by 
the central emission feature, with responses spanning along the wavelength direction of the spectral image in the Conv1 and becoming more concentrated around the central feature in the Conv2.

\section{Predictive uncertainty of CNN Score}\label{app:std_cnn}
Section~\ref{subsec:cl} refers to the potential model limitations, which primarily arise from the predictive uncertainty and the limited representativeness of the training set. 
This appendix evaluates the predictive uncertainty of the CNN score.

We conduct a Monte Carlo method by adding detector noise with a Gaussian distribution to each input spectral image. 
For each LAE candidate in the HETDEX COSMOS catalog, we randomly select up to $1000$ sources per \lya\ line S/N bin between 4.8 and 10.0, with a bin width of 0.4, excluding those flagged as AGNs. 
For each selected source, we generate $1000$ noise-added realizations of the spectral image and re-evaluate the CNN outputs. 
The standard deviation of the resulting CNN scores, $\sigma_{\mathrm{CNN\ Score}}$, is computed for each source. For each S/N bin, we then calculate the mean $\sigma_{\mathrm{CNN\ Score}}$ and the standard deviation of these means.

\begin{figure}[h]
\centering
    \includegraphics[trim=0 0 500 200, clip, width=0.45\textwidth]{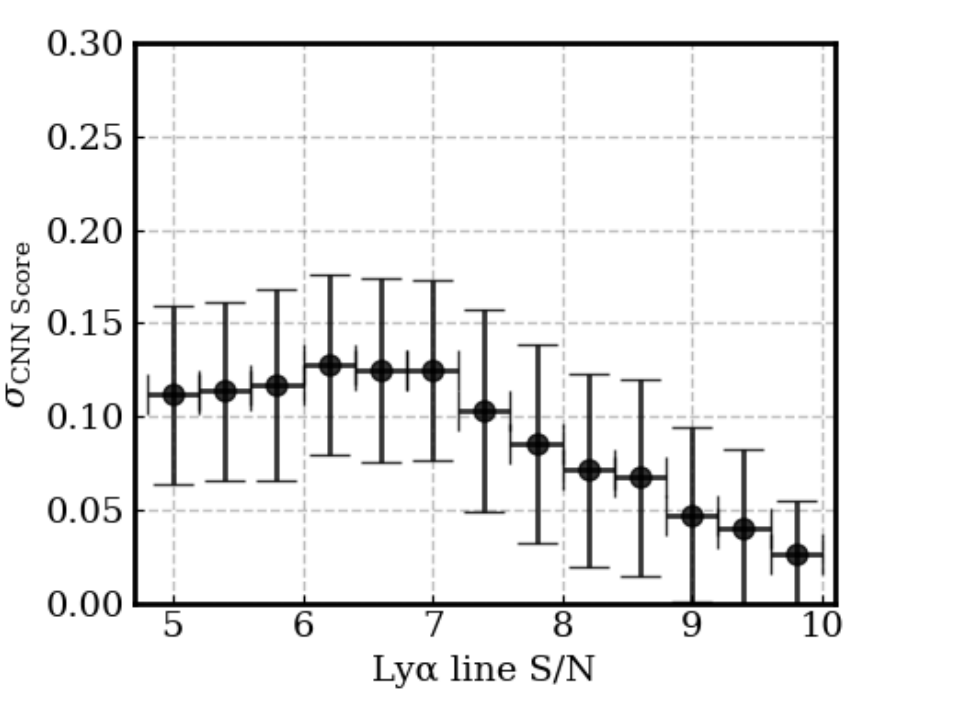}
    \caption{
    {\bf Standard deviation of CNN scores for LAE candidates as a function of S/N.} 
    Each point represents the mean $\sigma_{\mathrm{CNN\ Score}}$ in S/N bins of 0.4. 
    The error bars show the standard deviation of the mean values within each bin.
    }
    \label{fig:cnn_score_std}    
\end{figure}

Figure~\ref{fig:cnn_score_std} shows the mean $\sigma_{\mathrm{CNN\ Score}}$ as a function of 
S/N, with error bars representing the standard deviation of these mean estimates within each bin.
As predictions are consistent across all three folds, we present the results averaged over the three folds as the representative result.
The $\sigma_{\mathrm{CNN\ Score}}$ increases with decreasing S/N, indicating a predictive uncertainty of $\sigma_{\mathrm{CNN\ Score}}\lesssim0.17$ for S/N $\lesssim7$. 

\newpage

\bibliography{bibliography}
\bibliographystyle{aasjournal}

\end{document}